\documentclass[%
 aip,
rsi,%
 amsmath,amssymb,
reprint,%
]{revtex4-1}

\usepackage{amsmath,amsfonts,amssymb}
\usepackage{float}
\usepackage{gensymb}
\usepackage{mathrsfs}
\usepackage{subfigure}
\usepackage{url}
\usepackage{natbib}
\usepackage[usenames, dvipsnames]{color}
\usepackage{todonotes}
\usepackage{graphicx}
\usepackage{dcolumn}
\usepackage{bm}

\begin{document}


\title[Dynamics of two bouncing droplets]{Pilot-wave dynamics of two identical, in-phase bouncing droplets}

\author{Rahil N. Valani}
 \homepage{Electronic Mail: rahil.valani@monash.edu}
 \affiliation{School of Physics and Astronomy, Monash University, Victoria 3800, Australia}

\author{Anja C. Slim}
\affiliation{%
School of Mathematical Sciences, Monash University, Victoria 3800, Australia
}%
\affiliation{%
School of Earth, Atmosphere and the Environment, Monash University, Victoria 3800, Australia
}%

\date{25 June 2018}

\newcommand{\RV}[1]{{\protect \todo[inline, color =orange]{RV---#1}}}
\newcommand{\AS}[1]{{\protect \todo[inline, color =yellow]{AS---#1}}}
\newcommand{\JCM}[1]{{\protect \todo[inline, color =green]{JCM---#1}}}

\newcommand{\dis}{\ensuremath{\mathcal{D}}}
\newcommand{\dpw}{\ensuremath{\mathscr{D}}}
\newcommand{\upw}{\ensuremath{\mathscr{U}}}
\newcommand{\dcm}{\ensuremath{\mathfrak{D}}}
\newcommand{\ucm}{\ensuremath{\mathfrak{U}}}
\newcommand{\mm}{$\ensuremath{\text{mm}}$}
\newcommand{\cm}{$\ensuremath{\text{cm}}$}
\newcommand{\s}{$\ensuremath{\text{s}}$}
\newcommand{\Hz}{$\ensuremath{\text{Hz}}$}
\newcommand{\dd}{\ensuremath{d}}

\begin{abstract}
A droplet bouncing on the surface of a vibrating liquid bath can move horizontally guided by the wave it produces on impacting the bath.  The wave itself is modified by the environment, and thus the interactions of the moving droplet with the surroundings are mediated through the wave.  This forms an example of a pilot-wave system.  Taking the Oza--Rosales--Bush description for walking droplets as a theoretical pilot-wave model, we investigate the dynamics of two interacting identical, in-phase bouncing droplets theoretically and numerically.  A remarkably rich range of behaviors is encountered as a function of the two system parameters, the ratio of inertia to drag, $\kappa$, and the ratio of wave forcing to drag, $\beta$. The droplets typically travel together in a tightly bound pair, although they unbind when the wave forcing is large and inertia is small or inertia is moderately large and wave forcing is moderately small.  Bound pairs can exhibit a range of trajectories depending on parameter values, including straight lines, sub-diffusive random walks, and closed loops.  The droplets themselves may maintain their relative positions, oscillate towards and away from one another, or interchange positions regularly or chaotically as they travel.  We explore these regimes and others and the bifurcations between them through analytic and numerical linear stability analyses and through fully nonlinear numerical simulation.  
%
\end{abstract}

\maketitle

\begin{quotation}
A droplet of liquid can bounce on the surface of a bath of the same liquid indefinitely if the bath is experiencing vertical vibrations in an appropriate range of frequencies.  At each bounce, the droplet generates a decaying surface wave.  The slope of the wave at the point of the next bounce can impart a horizontal force on the droplet, leading to the droplet walking.  Since its discovery\cite{Couder2005} in 2005, this experimental system has excited significant research because it can exhibit quantum-like behavior.  The droplet is a particle that interacts with its environment through the wave it generates, \emph{i.e.}, it is a pilot-wave system.  In this paper we use a previously-published model~\cite{Oza2013} to explore a two-droplet pilot-wave system.  We find a remarkably wide range of behaviors, which we explore in detail.
\end{quotation}

 \section{\label{sec:Intro}Introduction}
In 2005, \citet{Couder2005} showed that if a bath of silicone oil is vibrated vertically with sinusoidal acceleration $\gamma \cos(2 \pi ft)$, with $\gamma$ the peak acceleration and $f$ the frequency, then a droplet of the same liquid as the bath can be made to bounce indefinitely on the oscillating surface provided $\gamma>\gamma_B$, with $\gamma_B$ the bouncing threshold. Just above the bouncing threshold, the droplet bounces at frequency $f$. As $\gamma$ increases, the droplet undergoes a series of bifurcations, and for $\gamma>\gamma_W>\gamma_B$, with $\gamma_W$ the walking threshold, a robust walking state develops for certain size droplets in which the droplet bounces at frequency $f/2$. This walking state emerges just below the Faraday instability threshold $\gamma_F$, above which the interface becomes unstable to standing Faraday waves of frequency $f/2$. Thus in this walking state, the droplet is a local exciter of Faraday waves whose decay time is a function of the `memory', the proximity to the Faraday threshold. At high memory (near the threshold), waves generated by the droplet in the distant past continue to affect it, and this hydrodynamic wave-particle system mimics several features thought to be intrinsic to the quantum realm. These include particle diffraction through single and double slit arrangements,\citep{Couder2006} orbital quantization in rotating frames\citep{Oza2014} and harmonic potentials,\citep{Perrard2014a,Perrard2014b} wavelike statistics in confined geometries\citep{PhysRevE.88.011001,Sáenz2017} and tunneling across submerged barriers.\citep{Eddi2009}

Exotic dynamics of a single walker have been observed both in experiments and numerical simulations in various situations. For example, \citet{PhysRevE.88.011001} experimentally showed that a single walker confined in a circular corral exhibits circular orbits at low memory with the emergence of more complex orbits such as wobbling circular orbits, drifting elliptical orbits and epicycles as the memory increases. At very high memory, the trajectory of the droplet becomes complex and chaotic with the emergence of a coherent wavelike statistical pattern. \citet{Perrard2014a,Perrard2014b} experimentally showed that a single walker in a harmonic potential at high memory undergoes complex motions such as ovals, lemniscates and trefoils in addition the circular orbits at low memories. They also showed that the constraints imposed on the dynamics of a walker by its pilot wave field results in a double quantization in the mean energy and angular momentum. Numerical simulations of a single walker in a rotating frame\citep{Oza2014a} and harmonic potential\citep{PhysRevFluids.2.113602} show similar quantized orbits. 

Studies of multiple walkers have been more limited. \citet{Protiere2006} reported the existence of bound states of two droplets such as parallel walkers, promenading pairs that oscillate towards and away from one-another while parallel walking, and tightly bound orbiting states. They also showed that multiple bouncing droplets self-organize into bound lattice structures. \citet{Borghesi2014} investigated the energy stored in the wave field of a promenading pair of walkers and related it to the interaction between the walkers. \citet{PhysRevE.78.036204} experimentally investigated  the dynamics of orbiting droplets of different sizes. Recently, \citet{PhysRevFluids.2.053601} and \citet{PhysRevFluids.3.013604} theoretically and experimentally investigated the orbiting and promenading states of identical droplets in detail and showed that the impact phase of walkers relative to the oscillations of the bath adjusts to stabilize the orbiting and promenading states.  Theoretical studies have focused on explaining these particular modes observed in experiments.  A study of the full dynamics of the two-droplet system as a function of the different parameters is lacking.

A number of theoretical models have been developed to describe the horizontal motion of a single droplet. The vertical dynamics of the droplets are fast compared to the horizontal and therefore in most models only the horizontal motion is considered and the walking droplet is assumed to continuously emit decaying Faraday waves strobed at the bouncing frequency. \citet{Protiere2006} developed the first model for the dynamics of a walking droplet that correctly predicts the bifurcation from bouncing to walking. Their model has a lumped drag term representing the average drag force on the droplets during each bounce, and a forcing term from the surface wave generated at the previous bounce.\citep{Oza2013} This approximation is only valid near the walking threshold where the waves are rapidly damped. \citet{Oza2013} proposed an improved model that includes forcing from surface waves generated at all previous bounces and also a lumped drag term describing the time-averaged drag force more quantitatively. In reality, each impact of a droplet on the bath generates a traveling wave, which is an order of magnitude faster than the walking speed of the droplet, and in its wake a standing wave remains that oscillates at the Faraday frequency.\citep{Eddi2009} In the Oza--Rosales--Bush description, only the standing wave part of the surface waves are modeled and these are approximated as zeroth-order Bessel functions.  For a single droplet, the transient wave does not interact with the droplet on subsequent bounces and may be safely ignored and the Bessel function for the wave field is observed to be a reasonably accurate representation within twice the Faraday wavelength.\citep{Milewski2015} The description further assumes that the impact phase of the droplet's bounce on the bath is constant, which limits the model's quantitative predictability to modes in which the phase remains constant.  \citet{Milewski2015} developed a more complete fluid model of pilot-wave hydrodynamics by coupling the vertical bouncing dynamics with a more accurate description of the weakly viscous quasi-potential wave generation and evolution. In this work, the problem was reduced to a two-dimensional version using Dirichlet-to-Neumann transformation in Fourier space. This model captures the transient wave and permits a more complete description of the standing wave field as well as allowing for an evolving bouncing phase, at the expense of solving for wave generation in the bath on the time-scale of a single bounce.

Stroboscopic models for multiple droplets are still in development.  In walker-walker interactions, neglecting the transient wave and assuming a simplified wave structure are reasonable within two Faraday wavelengths, but the assumption of constant phase breaks down and thus far this has only been addressed using an empirical fix for the particular system being considered.\citep{PhysRevFluids.2.053601,PhysRevFluids.3.013604}




Here we take the Oza--Rosales--Bush description\citep{Oza2013} as a theoretical pilot-wave model and explore the behaviors predicted for a simple extension to dynamics of two identical, in-phase bouncing droplets. We find parallel walkers and promenading pairs as well as a rich array of more exotic dynamics such as regularly and chaotically switching walkers, wandering walkers and intriguing closed-loop trajectories in regions of parameter space where wave forcing and/or inertia play a significant role.

In  Section~\ref{Fm}, we describe the equations of motion for two droplets. In Section~\ref{ps}, we give an overview of the range of behaviors possible across parameter space and describe these behaviors in more detail in Sections~\ref{SS} to~\ref{ed}.  We conclude with a discussion in Section~\ref{sec:concl}.

\begin{figure*} 
\centering
\includegraphics[width=15cm]{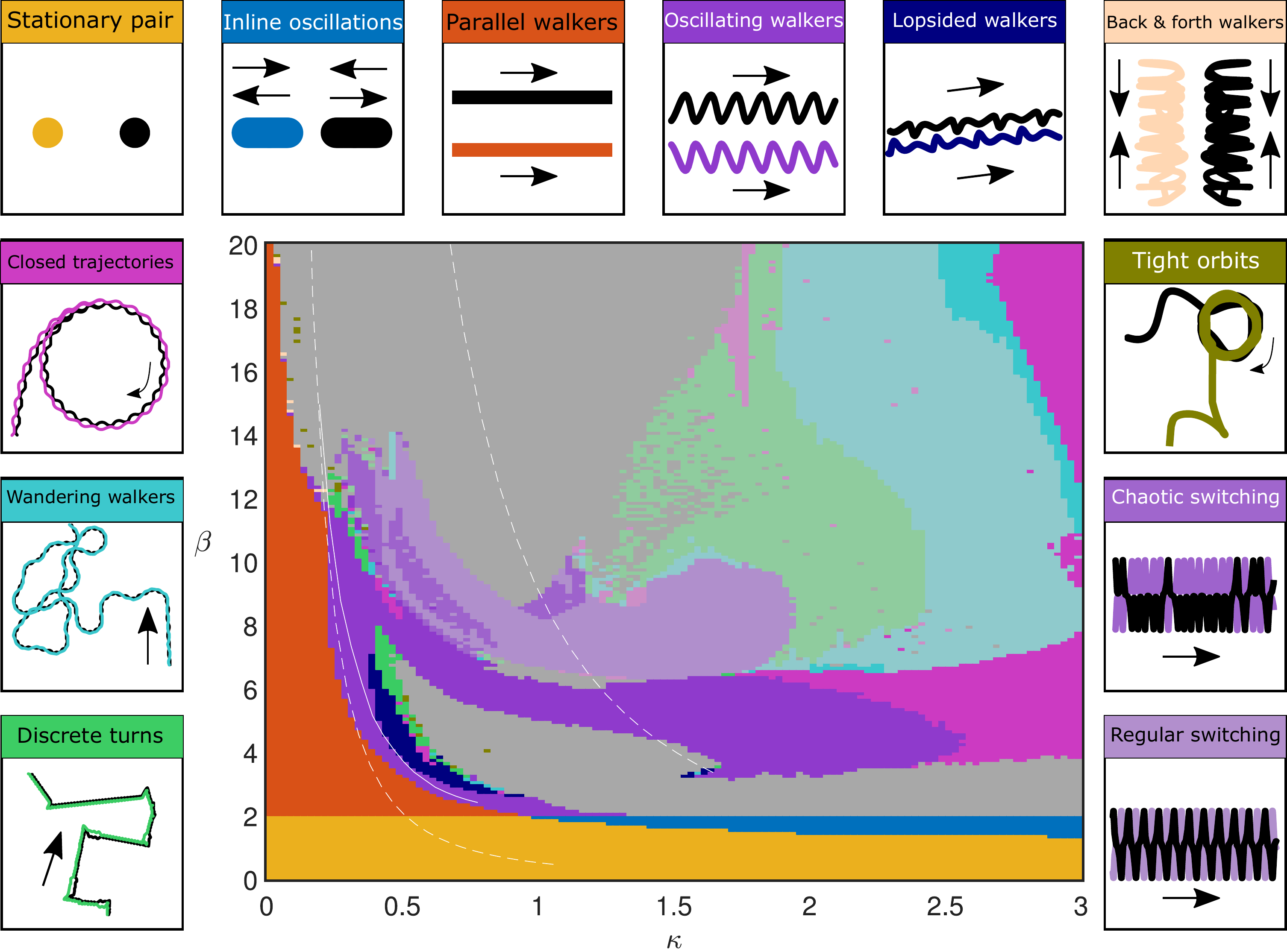}
\caption{Behavior observed in the $\beta$-$\kappa$ parameter space at $t=1000$ from simulations initiated at $t=0$ as parallel walkers with noise. We explore the parameter space region $0<\kappa \leq3$ and $0< \beta\leq 20$ with resolution $\Delta \kappa=0.025$ and $\Delta \beta=0.1$. In gray regions the droplets have become unbound. The colored regions correspond to the various states depicted in the surrounding trajectory plots.  For oscillating walkers (purple), discrete-turning walkers (green), wandering walkers (cyan), and closed trajectories (pink), the darker shaded regions have non-switching oscillating walkers and the lighter regions have regularly switching walkers.  The intermediate shade for oscillating walkers exhibit some form of chaotic switching. We note that the back-and-forth walkers may only be transient behaviour as we find that in our simulations, some of the back-and-forth walkers ultimately either settle into a tight orbit or become unbound.  The region between the faint dashed white curves indicates where existing experimental setups may be able to perform experiments (see also \citet{ValaniHOM} Fig.~3). The solid white curve is the transect along which \citet{PhysRevFluids.3.013604} observed oscillating walkers for in-phase bouncing droplets at the closest inter-droplet distance $\dpw_1$.}
\label{fig:parameter_space}
\end{figure*}

\section{Formulation}\label{Fm}
Consider two identical droplets bouncing in-phase on the surface of a bath oscillating vertically at frequency $f$.  The dimensionless positions of the droplets in the horizontal plane are $\mathbf{r}_1=(x_1,y_1)$ and $\mathbf{r}_2=(x_2,y_2)$.  We describe their horizontal motion by the pair of integro-differential equations
\begin{align}\label{eq_1}
\kappa\ddot{\mathbf{r}}_{i}+\dot{\mathbf{r}}_{i} = -\beta \left.\nabla h(\mathbf{r},t)\right|_{\mathbf{r}=\mathbf{r}_i(t)}
\end{align}
for $i=1$, $2$, where the dimensionless height of the interface
 \begin{align}\label{h_eq}
{h}(\mathbf{r},t) &= \int_{-\infty}^{t}\text{J}_0(|\mathbf{r} - \mathbf{r}_{1}(s)|)\text{e}^{-(t-s)} \text{d}s \notag\\
& + \int_{-\infty}^{t}{\text{J}_0(|\mathbf{r} - \mathbf{r}_{2}(s)|)}\text{e}^{-(t-s)} \text{d}s
\end{align}
and dots indicate differentiation with respect to dimensionless time $t$.
This is the direct extension of the single-droplet model developed by \citet{Oza2013} to a two-droplet system. The left hand side of the equation comprises an inertial term $\kappa\ddot{\mathbf{r}}_{i}$ and an effective drag term $\dot{\mathbf{r}}_{i}$.  The right hand side of the equation captures the forcing of the droplets by the waves they have generated.  Each impact generates a wave modeled as an axisymmetric Bessel function $\text{J}_0(\mathbf{r})$ centered at the point of impact and decaying exponentially in time. Since this model takes into account the waves generated from all the previous impacts, the shape of the interface is calculated through integration of waves generated from all the previous bounces of both the droplets. At each impact, the droplet receives a horizontal kick proportional to the gradient of the interface at that point. The dimensionless parameters $\kappa$ and $\beta$ follow directly from \citet{Oza2013} and are referred to as the dimensionless mass and the memory force coefficient respectively.  They may be usefully interpreted as the ratios of inertia to drag and wave forcing to drag respectively.  Thus for small $\kappa$, the droplets' motion responds effectively immediately to the wave forcing.  For large $\kappa$, it responds more slowly and a more sustained forcing is required to modify the motion.  In such regimes, the droplets are likely to overshoot their equilibria and oscillations are likely.  This model can be extended to two identical out-of-phase bouncing droplets by appropriately switching the signs of the forcing term on the right hand side of equation \eqref{eq_1} for the two droplets. Moreover, two droplets of different sizes can be obtained by using different $\kappa$ and $\beta$ for each droplet. In this study, we only focus on the dynamics of two identical, in-phase bouncing droplets.

For the details of the non-dimensionalisation, we refer the reader to \citet{Oza2013}.  However, we note that the length scale is chosen such that the Faraday wavelength is $2\pi$ and the time scale is such that the Faraday period is $1/M_e$ where $M_e$ is the memory parameter which represents the proximity to the Faraday threshold.\citep{Oza2013} For the parameter space under consideration, this memory typically varies in the range $1\lesssim M_e \lesssim 20$. 

We numerically integrate equations \eqref{eq_1} using a modified Euler method with a dimensionless time step $\Delta t=2^{-6}$ unless stated otherwise. In this method, the new position is calculated from the current velocity using a Forward Euler step but the new velocity is calculated from the new position using a Backward Euler step. The details of the numerical method are provided in Appendix \ref{numerics}.  

\section{Parameter space description}\label{ps}


We begin with a summary of the rich dynamics observed on varying $\beta$ and $\kappa$.  Fig.~\ref{fig:parameter_space} shows the behavior at $t=1000$ for droplets initiated at $t=0$ as parallel walkers with noise. Specifically, the initial positions were taken as $\mathbf{r}_1=(0,0)$ and $\mathbf{r}_2=(\dpw_1(\beta),0)$ and the initial velocities as $\dot{\mathbf{r}}_1=(\delta_1,\upw_1(\beta)+\delta_2)$ and $\dot{\mathbf{r}}_2=(\delta_3,\upw_1(\beta)+\delta_4)$, where $\dpw_1(\beta)$ and $\upw_1(\beta)$ are the distance between the two droplets and the velocity of each droplet in the parallel walking state (described in Section~\ref{PW}), and each $\delta_i$ is a random perturbation uniformly selected between $-0.1$ and $0.1$. For $t<0$, the droplets were assumed to be in the unperturbed parallel walking state.  



For $\kappa<1$ where drag exceeds inertia, a bifurcation from stationary states (yellow) to walking states occurs at $\beta=2$, as for single droplets.\citep{Oza2013}  For $\kappa>1$ where inertia exceeds drag, the droplets are stationary for very small wave forcing $\beta$, before starting to oscillate towards and away from one another about fixed positions for $\beta$ in a region below and very slightly above $2$.  We term this latter behavior \emph{inline oscillations} (blue). For $\beta>2$, we observe a variety of walking motions. For $\kappa<1$ and moderate $\beta$, the droplets perform a parallel walk at constant velocity. These states have been observed experimentally and are referred to as \emph{parallel walkers} (red).\citep{protière_boudaoud_couder_2006} For larger $\beta$, the droplets oscillate, predominantly towards and away from one another, while walking. These states have also been observed experimentally and have been referred to as promenading pairs.\citep{Borghesi2014}  We refer to them as \emph{oscillating walkers} (purple) to simplify classification. Upon further increasing $\beta$, these oscillating walkers tend to unbind.  More exotic dynamics such as \emph{lopsided walkers} (navy blue), \emph{regular switching walkers} (purple - light shade), \emph{chaotic switching walkers} (purple - intermediate shade), \emph{back-and-forth walkers} (beige), \emph{discrete-turning walkers} (green), \emph{continuously turning walkers} (sky blue) and \emph{closed trajectories} (pink) are observed for larger $\beta$ and $\kappa$. These various states are explored in the next sections: stationary states in Section~\ref{SS}, inline oscillations in Section~\ref{inlineosc}, parallel walkers in Section~\ref{PW}, oscillating walkers in Section~\ref{sec:ow} and more exotic, wandering states in Section~\ref{ed}.  Despite the initial conditions being those of parallel walkers, we also very occasionally observe the droplets binding into tight orbits for large $\beta$ and very small $\kappa$. We refer the reader to \citet{PhysRevFluids.2.053601} for more details on this state.  

\section{Stationary States}\label{SS}

We begin by exploring stationary states.  Consider two droplets a distance $d$ apart. We will look for equilibrium states of the system such that the droplets remain stationary at this distance. Substituting $\mathbf{r}_{1}=(0,0)$ and $\mathbf{r}_{2}=(d,0)$ into \eqref{eq_1}, we obtain the constraint
\begin{equation}
\text{J}_1(d)=0. 
\end {equation}
We denote the discrete solutions of this equation by $d=\dis_n$, where $\dis_n$ is the $n$th zero of the Bessel function $\text{J}_1(\cdot)$. At these equilibrium distances, the second droplet sits either at a trough (odd $n$) or a crest (even $n$) of the wave field generated by the first droplet $(\text{J}_0'(d)=-\text{J}_1(d)=0)$. We will focus on the first four distances $\dis_1\approx 3.83$, $\dis_2\approx 7.02$, $\dis_3\approx10.17$ and $\dis_4\approx13.32$. 

\subsection{Linear stability analysis}\label{sec:ss_stability}
To investigate the stability of these stationary states, we consider a general perturbation to the droplets: $\mathbf{r}_{1}=(0,0)+\epsilon({x_{11}}(t),{y_{11}}(t))$ and $\mathbf{r}_{2}=(d,0)+\epsilon({x_{21}}(t),{y_{21}}(t))$. Substituting these forms into \eqref{eq_1} and linearizing the resulting equations, we obtain the matrix equation
\begin{equation}
 \begin{bmatrix}
   \dot{\mathbf{X}}_{1} \\
    \dot{\mathbf{Y}}_{1}\\
   \dot{\mathbf{X}}_{2} \\
\dot{\mathbf{Y}}_{2} \\
  \end{bmatrix}=
  \begin{bmatrix}
    \mathbf{\Omega} & \boldsymbol{\mathcal{O}} & \boldsymbol{\chi} & \boldsymbol{\mathcal{O}} \\
    \boldsymbol{\mathcal{O}} & \mathbf{\Theta} & \boldsymbol{\mathcal{O}} & \boldsymbol{\mathcal{O}} \\
\boldsymbol{\chi} & \boldsymbol{\mathcal{O}} & \mathbf{\Omega} & \boldsymbol{\mathcal{O}} \\
\boldsymbol{\mathcal{O}} & \boldsymbol{\mathcal{O}} & \boldsymbol{\mathcal{O}} & \mathbf{\Theta} \\
  \end{bmatrix}
\begin{bmatrix}
   \mathbf{{X}}_{1} \\
   \mathbf{ {Y}}_{1}\\
   \mathbf{{X}}_{2} \\
\mathbf{{Y}}_{2} \\
  \end{bmatrix}
  \label{SS_matrix}
\end {equation}
where
\begin{equation*}
 \mathbf{{X}}_{i} =
  \begin{bmatrix}
    {x}_{i1} \\
    \dot{x}_{i1} \\ 
    {X}_{i1} \\
  \end{bmatrix},
\quad
 \mathbf{{Y}}_{i} =
  \begin{bmatrix}
    {y}_{i1} \\
    \dot{y}_{i1}\\
{Y}_{i1} \\
  \end{bmatrix}
\end {equation*}
for $i=1$, $2$; 
\begin{gather*}
 \mathbf{\Omega} =
  \frac{1}{2 \kappa}\begin{bmatrix}
    0 & 2\kappa & 0 \\
    \beta\left( 1+2\text{J}_1'(d) \right) & -2 & -\beta\\
2\kappa & 0 & -2\kappa \\
  \end{bmatrix},\\
  \boldsymbol{\chi} =
 \frac{1}{\kappa} \begin{bmatrix}
    0 & 0 & 0 \\
    0 & 0 & -\beta \text{J}_1'(d)\\
0 & 0 & 0 \\
  \end{bmatrix}, \quad
\mathbf{\Theta} =
  \frac{1}{2 \kappa}\begin{bmatrix}
    0 & 2\kappa & 0 \\
    \beta & -2 & -\beta\\
2\kappa & 0 & -2\kappa \\
  \end{bmatrix},
\end {gather*}
and $\boldsymbol{\mathcal{O}}$ is the $3\times3$ Zero matrix. Derived variables $X_{i1}$ and $Y_{i1}$ are given by
\begin{gather*}
X_{i1}=\int_{-\infty}^{t}x_{i1}(s)\text{e}^{-(t-s)} \text{d}s, \text{ }
Y_{i1}=\int_{-\infty}^{t}y_{i1}(s)\text{e}^{-(t-s)} \text{d}s.
\end{gather*}

\begin{figure*}
\subfigure{\raisebox{11mm}{\includegraphics[width=2.2cm]{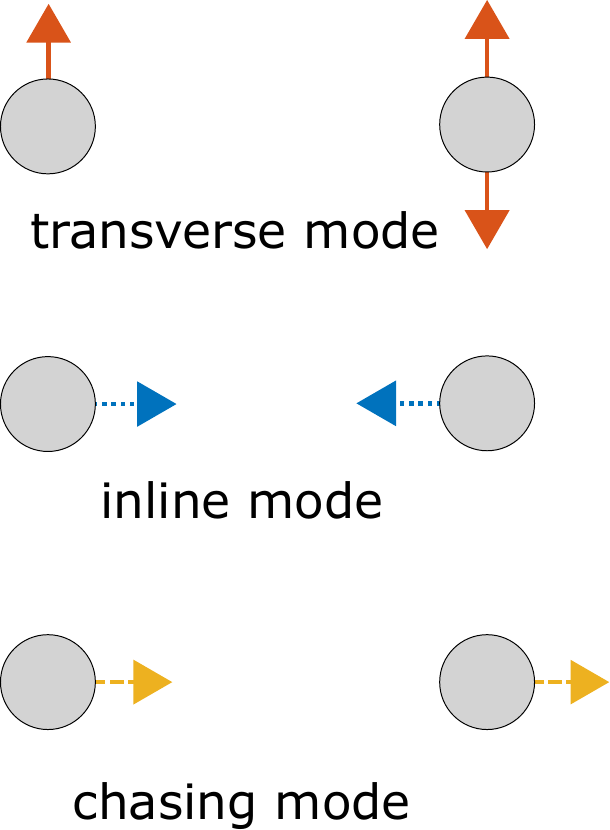}}}
\hspace{-0.15cm}\subfigure{\includegraphics[height=4.8cm]{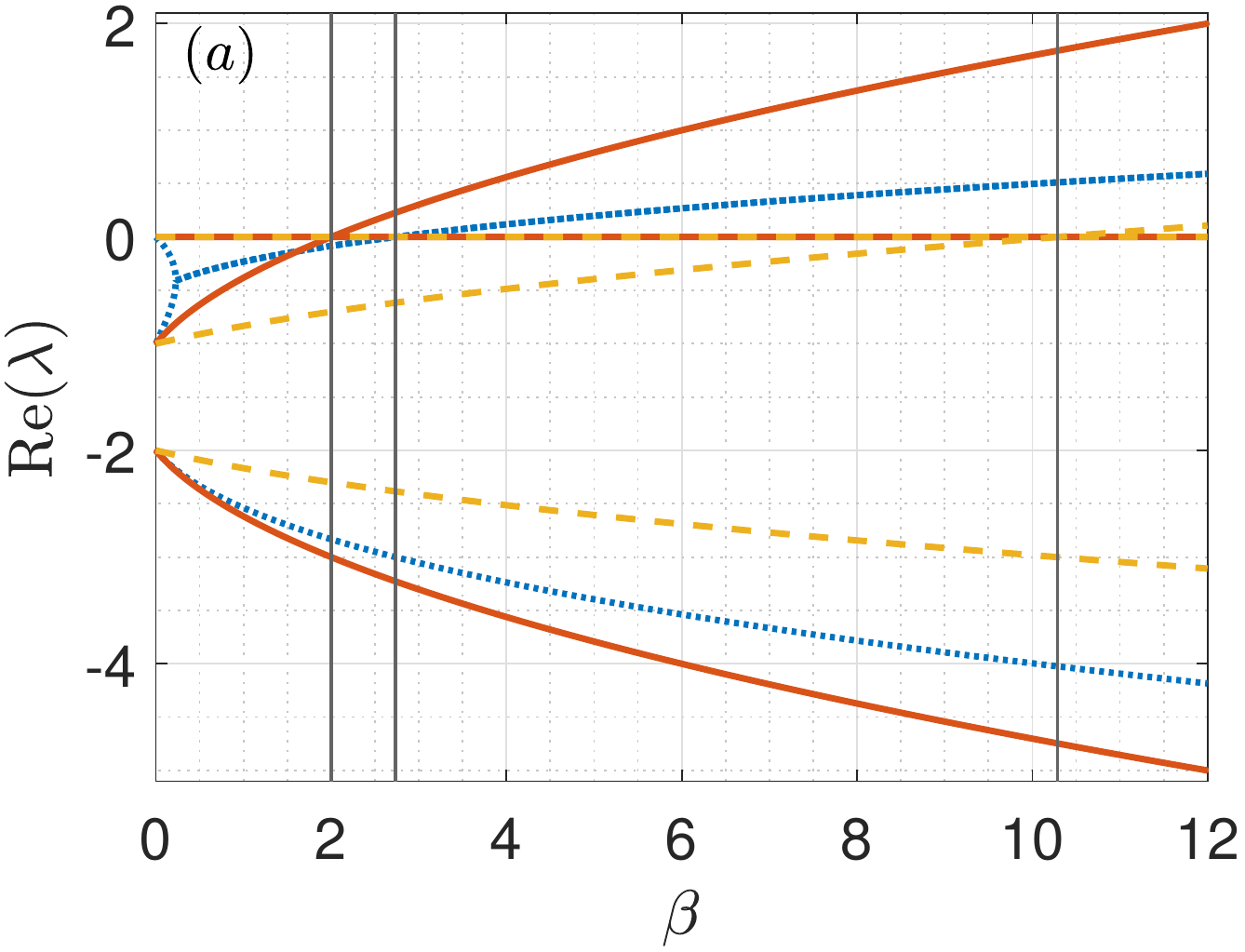}}
\subfigure{\includegraphics[height=4.8cm]{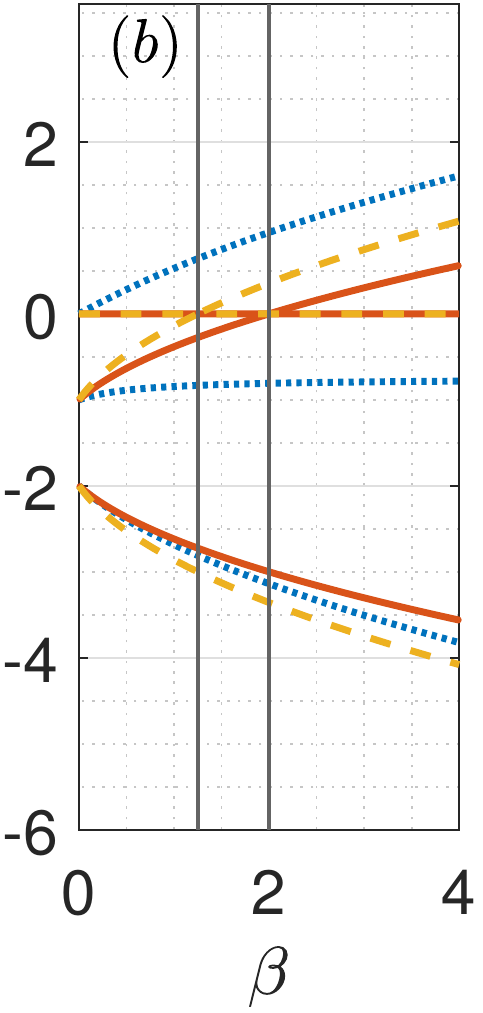}}
\hspace{-0.15cm}\subfigure{\includegraphics[height=5.2cm]{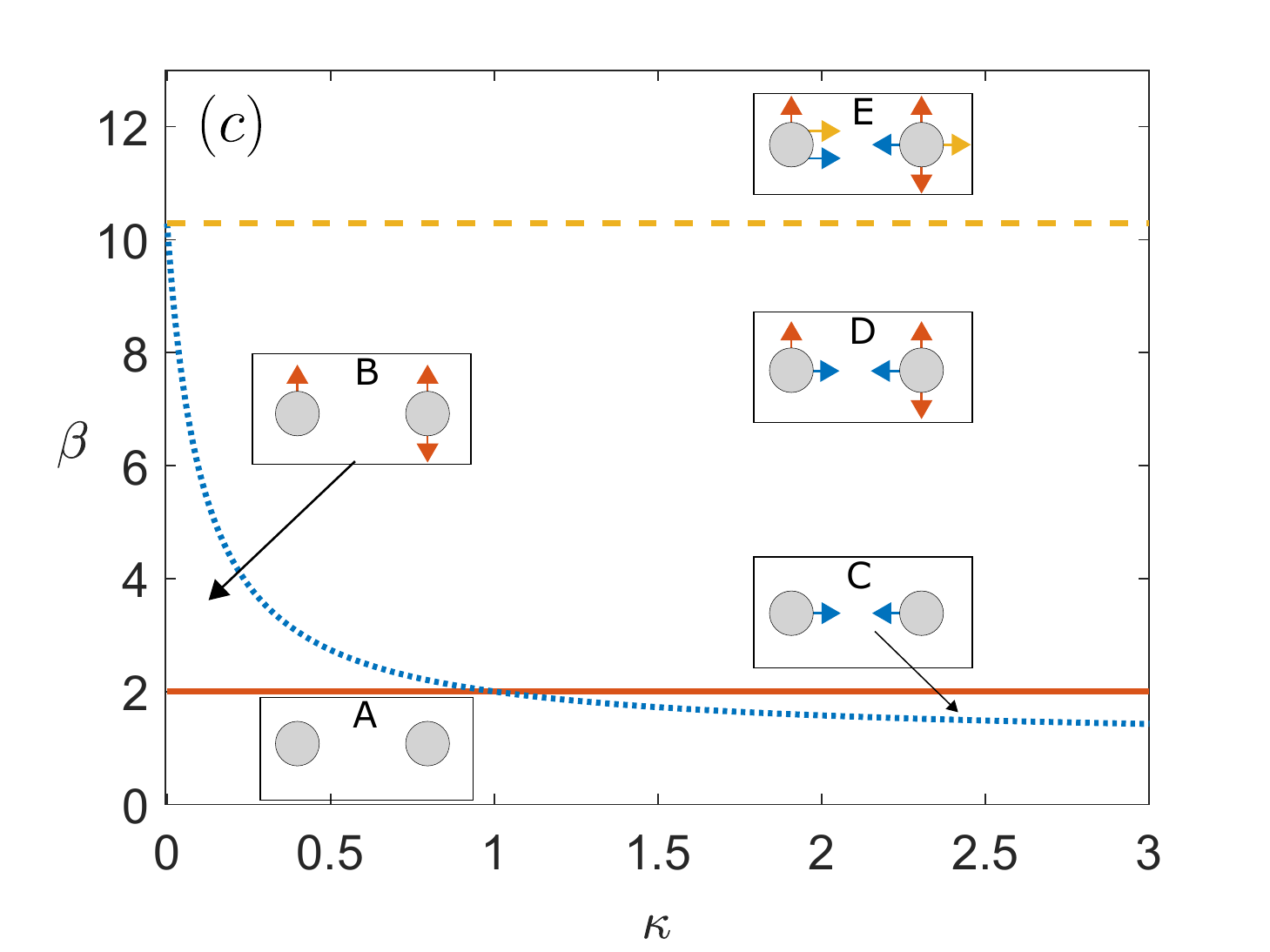}}
\caption{Stationary pairs: Linear growth rates of perturbations $Re(\lambda)$ as a function of memory parameter $\beta$ for droplets (a) the first stationary distance $d=\dis_1$ apart and (b) the second stationary distance $d=\dis_2$ apart.  Perturbation modes are distinguished as transverse (red, solid curves), inline (blue, dotted curves) or chasing (yellow, dashed curves). The vertical lines shows the $\beta$ values at which the eigenvalues cross Re$(\lambda)=0$.  The dimensionless mass $\kappa = 0.5$. (c) Stability diagram in the $\beta$-$\kappa$ parameter space for the first stationary distance $d=\dis_1$. Region A is stable to any small perturbation. Regions B and C are unstable to transverse and inline perturbations respectively. Region D is unstable to both inline and transverse perturbations while Region E is also unstable to chasing modes perturbations. }
\label{fig:eig_ss}
\end{figure*}


The solutions of \eqref{SS_matrix} are proportional to $\text{e}^{\lambda t}$, with the complex growth rates $\lambda$ given by the eigenvalues of the right-hand-side matrix. The characteristic polynomial of this matrix factorizes in a convenient manner as
\begin{equation*}
\det(\lambda \boldsymbol{I} - \mathbf{\Theta})^2 \det(\lambda \boldsymbol{I} -\mathbf{\Omega}-\boldsymbol{\chi})\det(\lambda \boldsymbol{I} - \mathbf{\Omega}+\boldsymbol{\chi})=0,
\end{equation*}
where each of the sub-determinants corresponds to a distinct eigenmode of the system. Thus
\begin{align*}
\det(\lambda \boldsymbol{I} - \mathbf{\Theta})
&=\lambda^3 + \frac{\kappa+1}{\kappa} \lambda^2 - \frac{\beta-2}{2 \kappa}\lambda
\end{align*}
is the characteristic polynomial corresponding to perturbations perpendicular to the line joining the droplets called the \emph{transverse mode},
\begin{gather}
F_{i}(\lambda):= \det(\lambda \boldsymbol{I} - \mathbf{\Omega}+\boldsymbol{\chi})
= \lambda^3 + \frac{\kappa+1}{\kappa}\lambda^2 \notag \\
- \frac{\beta (2 \text{J}_1'(d)+1)-2}{2 \kappa}\lambda -\frac{2 \text{J}_1'(d) \beta}{\kappa} \label{eq:inline}
\end{gather}
corresponds to inline perturbations of the droplets towards or away from one other called the \emph{inline mode} and
\begin{equation*}
\det(\lambda \boldsymbol{I} - \mathbf{\Omega}-\boldsymbol{\chi})
=\lambda^3 + \frac{\kappa+1}{\kappa}\lambda^2
-\frac{\beta(2 \text{J}_1'(d)+1)-2}{2 \kappa}\lambda
\end{equation*}
corresponds to inline perturbations of the droplets in the same direction called the \emph{chasing mode}.

Fig.~\ref{fig:eig_ss}(a,b) shows the growth rates as a function of the memory force parameter $\beta$ for the two smallest stationary distances. For $d=\dis_1$, when $\beta<2$, the real part of all the non-trivial eigenvalues are negative indicating that the two-droplet system is stable for general small perturbations. When $\beta\geq 2$, an eigenvalue for each distinct mode becomes positive at different $\beta$ values. Note that there are also two zero eigenvalues, which correspond to invariants of the equilibrium state.

Transverse perturbations become unstable at $\beta=2$ independent of $\kappa$. 
This bifurcation value is identical to that for a single droplet's bouncing-to-walking transition.\cite{Oza2013}  This is not a coincidence: for transverse perturbations, the order-$\epsilon$ forcing to each droplet arises only from the droplet's own wave field while the contribution from the other droplet's wave field is of higher order. Thus the linearized equations for the two droplets decouple and reduce to those of a single droplet. The eigenvalues of the transverse mode are purely real. At the onset of instability, parallel walkers emerge if the droplets are perturbed in the same transverse direction, while orbiting states emerge if the droplets are perturbed in the opposite transverse direction. The parallel walking state will be explored in Section~\ref{PW}. 

For the inline mode, a pair of complex conjugate eigenvalues become unstable at 
\begin{equation}\label{inbif}
\beta^i_n=\left(\frac{1}{2}-\text{J}_1'(\dis_n)\left(\frac{\kappa-1}{\kappa+1}\right)\right)^{-1}.
\end{equation}
At the onset of instability, the droplets oscillate towards and away from one another with angular frequency 
\begin{equation*}\label{inline1}
\omega_n=\sqrt{\frac{1}{2 \kappa}(2-\beta(2 \text{J}_1'(\dis_n) + 1))}.
\end{equation*}
These oscillations, termed inline oscillations, are discussed in Section~\ref{inlineosc}.  

For the chasing mode, the eigenvalues are purely real and an eigenvalue become unstable at
\begin{equation*}
\beta^c_n  =  \frac{2}{2\text{J}_1'(\dis_n)+1}
\end{equation*}
independent of $\kappa$. For $d=\dis_1$, this corresponds to $\beta^c_1 \approx10.29$. In this mode, the droplets walk one behind another in the same direction at a constant speed. These chasers are explored briefly in Appendix~\ref{chase}. 

Fig.~\ref{fig:eig_ss}(c) summarizes the linear stability of stationary states at $d=\dis_1$. There are regions where only the walking or the inline oscillating mode is unstable while the chasing mode bifurcation only takes place where both inline and transverse modes are unstable. The bifurcations from stationary states to parallel walking and stationary states to inline oscillations match with the states observed numerically in Fig.~\ref{fig:parameter_space}.

From Fig.~\ref{fig:eig_ss}(b) it is clear that at $d=\dis_2$, one eigenvalue for inline perturbations always has positive real part and therefore any perturbations will drive the system away from the stationary state. Considering only the eigenvalues corresponding to inline perturbations given in equation \eqref{eq:inline} and by invoking Descartes' rule of sign, we can deduce the existence of one positive root of this cubic equation. Thus the equilibrium distances $\dis_{2n}$ are always unstable to inline perturbations, as expected since one droplet is sitting on the crest of the other's wave field at these distances and small perturbations will result in kicks away from the equilibrium. 


\section{Inline oscillations}\label{inlineosc}

\begin{figure*}
\centering
\includegraphics[width=2\columnwidth]{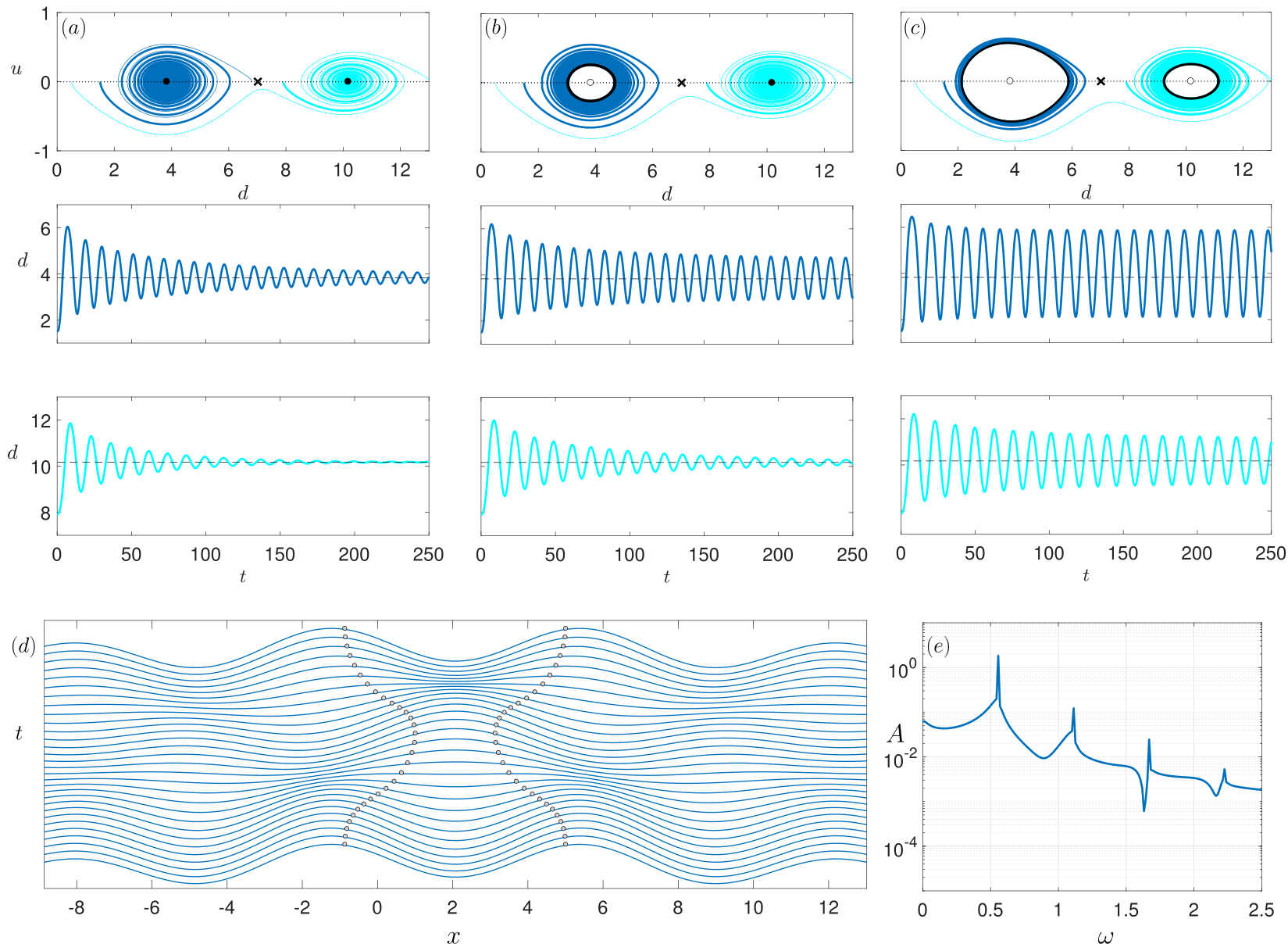}
\caption{Inline oscillations: $u$-$d$ phase plane at $\kappa=2$ and (a) $\beta=1.5$, (b) $\beta=1.6$ and (c) $\beta=1.75$. (a) At $\beta=1.5$, both $d=\dis_1$ and $d=\dis_3$ are stable spirals (filled black circles) while $d=\dis_2$ is a saddle (black cross). (b) At $\beta$=1.6, $d=\dis_3$ is still a stable spiral while at $d=\dis_1$, an unstable spiral (empty black circles) has emerged with an enclosing limit cycle. (c) At $\beta$=1.75, limit cycles exist at both $\dis_1$ and $\dis_3$. The two panels below the phase plane plot shows the distance between the droplets as a function of time for the thick solid line trajectories in the phase plane. For $\beta$=1.75, (d) shows cross-sections of the wave field generated by the droplets and droplet positions at five different instants over one period of the limit cycle at $\dis_1$ and (e) shows the FFT of the distance between the two droplets indicating that the oscillations are dominated by a single frequency.
}

\label{phase_inline}
\end{figure*}


In a sliver of parameter space with inertia exceeding drag, $\kappa>1$, and small wave forcing $\beta\lesssim 2$, inline oscillations are observed with droplets oscillating towards and away from one another (see Fig.~\ref{fig:parameter_space}). 
Here we explore the nature of the oscillations. 

The phase space for the one-dimensional inline motion of the droplets is two dimensional with the velocity $u(t)$ of the first drop and the distance $d(t)$ between the droplets sufficient to fully describe the system.  The evolution of the phase-space portrait with increasing wave forcing $\beta$ at fixed $\kappa=2$ is shown in Fig.~\ref{phase_inline}. Note that, from equation \eqref{inbif}, the $\beta$ value at which different distances $\dis_{n}$ with $n$ odd become unstable are a function of $\kappa$.  For $0<\kappa<1$, the cascade of instability goes from larger to smaller distances as $\beta$ increases, while for $\kappa>1$ it goes from smaller to larger distances. For $\kappa=1$, all the distances become unstable at the same value $\beta^i_n=2$.  For $\kappa=2$, the onset of inline oscillations occurs at $\beta^i_1 =1.577$ for $\dis_1$ and $\beta^i_3 =1.715$ for $\dis_3$. Thus at $\beta=1.5$ (Fig. \ref{phase_inline}(a)) there are stable spirals at $\dis_1$ and $\dis_3$ and a saddle at the unstable distance $\dis_2$.  If the droplets are perturbed inline when placed near a distance $\dis_1$ or $\dis_3$ apart, the oscillations will decay and the droplets will settle back into the stationary distance. As the parameter $\beta$ is increased beyond $\beta^i_1$, the stable spiral at $\dis_1$ appears to undergo an apparent supercritical Hopf bifurcation and changes into an unstable spiral with an encompassing limit cycle (Fig.~\ref{phase_inline}(b)).  Now the droplets either perform limit cycle oscillations corresponding to motion towards and away from one another around $\dis_1$ or settle into the second stable distance $\dis_3$. On further increasing $\beta$ beyond $\beta^i_3$, the stable spiral at $\dis_3$ also undergoes a supercritical Hopf bifurcation as shown for $\beta=1.75$ in Fig.~\ref{phase_inline}(c). Eventually, as $\beta$ is increased beyond $1.90$, the limit cycle at $\dis_1$ vanishes in an apparent homoclinic bifurcation, followed by that at $\dis_3$ at $\beta=2.01$.  

In simulations, inline oscillations are observed with rapidly increasing separation $\dis_n$ as $\beta$ is increased for fixed $\kappa$.  The droplets unbind in the simulations when $\beta \approx 2.1$.

Fig.~\ref{phase_inline}(d) shows a representative example of the positions of the droplets along with cross-sections of their wave field for one oscillation cycle. When the droplets are at their maximum separation, the wave field gradient ensures a kick towards each other. As the droplets travel towards each other, they pass their mean distance and reach a minimum separation with wave field gradient such that the droplets receive a kick away from each other. In this way, the droplets oscillate towards and away from one another. Note that the oscillations are dominated by a single frequency (Fig.~\ref{phase_inline}(e)) and a single Fourier mode expansion approximates the oscillations reasonably well near the bifurcation.

\section{Parallel walking}\label{PW}

\begin{figure}
\hspace*{-1.2cm} 
\centering
\subfigure{%
\includegraphics[width=6.9cm]{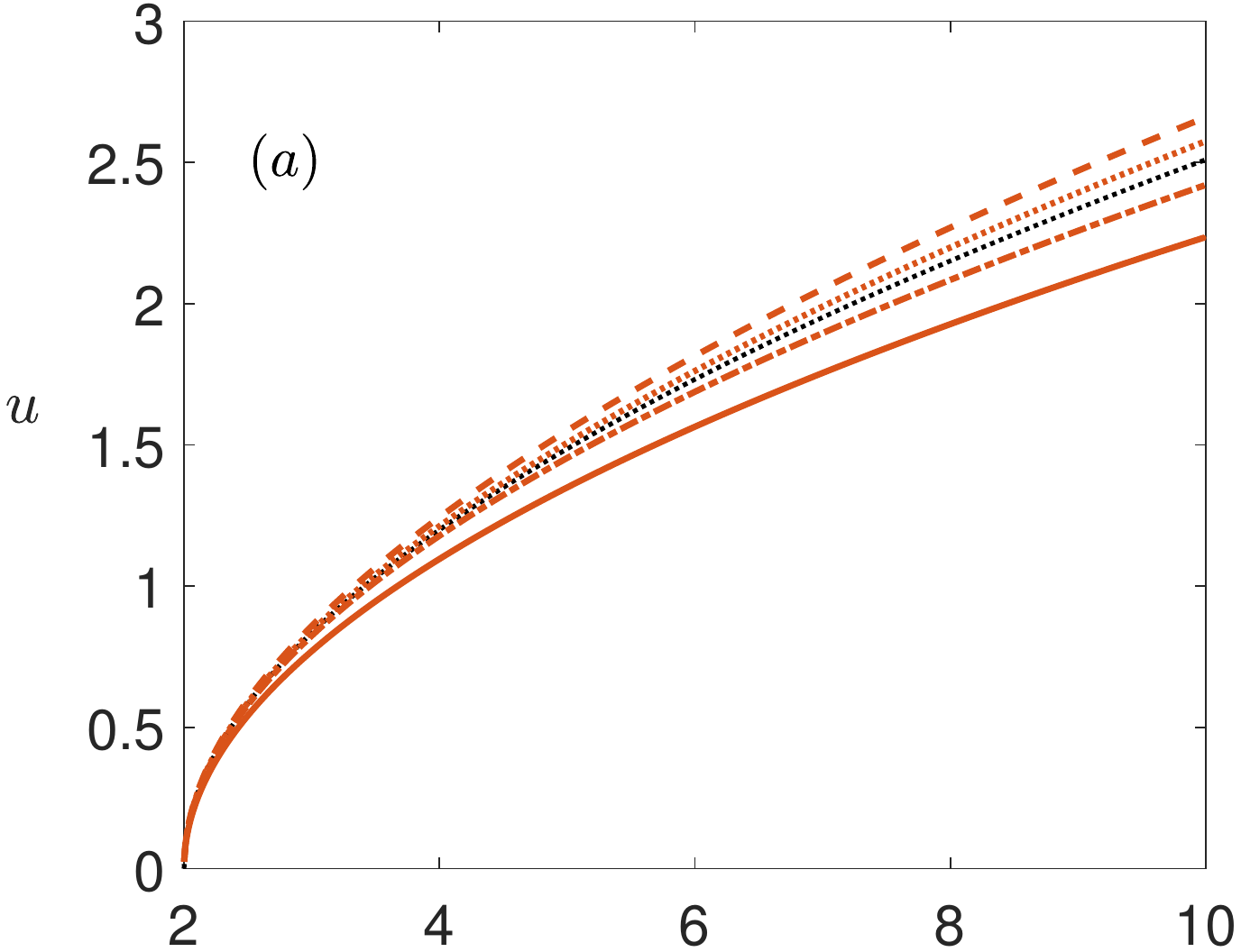}}
\quad 
\hspace*{-1cm} 
\subfigure{%
\includegraphics[width=6.77cm]{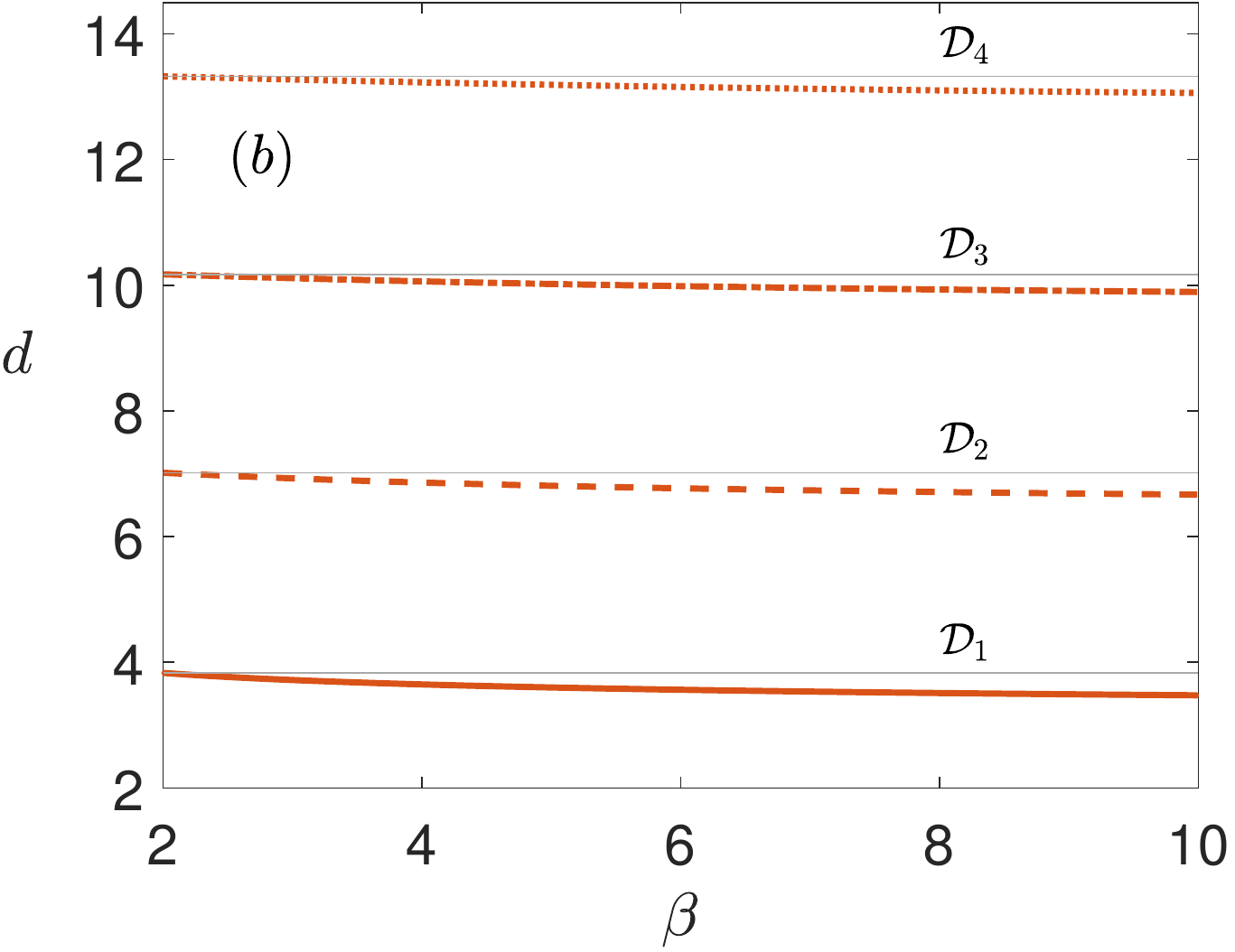}}
\caption{Parallel walkers: First four equilibrium (a) walking speeds $u=\upw_n(\beta)$ and (b) separations $d=\dpw_n(\beta)$ as a function of $\beta$. Based on the linear stability analysis, the distances $\dpw_1(\beta)$ and $\dpw_3(\beta)$ are stable while $\dpw_2(\beta)$ and $\dpw_4(\beta)$ are unstable. The black curve in (a) represents the solution for a single walker. The black lines in (b) are the stationary state equilibrium distances.}
\label{PW_eq}
\end{figure}

For $\kappa<1$ and $\beta>2$, a parallel walking state emerges in which the droplets walk at constant speed in the direction perpendicular to the line joining them. Consider two such droplets moving at constant speed $u$ and separated by a distance $d$. By substituting $\mathbf{r}_{1}=(0,ut)$ and $\mathbf{r}_{2}=(d,ut)$ in equation \eqref{eq_1}, we arrive at the pair of integral equations
\begin{gather*} \label{eq:pw}
 \int_{0}^{\infty}\frac{\text{J}_1(\sqrt{u^2z^2 + d^2})}{\sqrt{u^2{z^2} + d^2}}\text{e}^{-z} \text{d}z=0 \
\end{gather*}
and
\begin{gather*}
\frac{u}{\beta} = \frac{\sqrt{1+u^2}-1}{u\sqrt{1+u^2}}
+  \int_{0}^{\infty}\frac{{uz}\,\text{e}^{-z}}{\sqrt{u^2{z^2} + d^2}}\text{J}_1(\sqrt{u^2z^2 + d^2}) \text{d}z.
\end{gather*}
These can be solved numerically and have infinitely many solutions $u=\upw_n(\beta)$, $d=\dpw_n(\beta)$ for $n \in \mathbb{N}$, which are functions of $\beta$ but independent of $\kappa$. The first four such solutions are shown in Fig.~\ref{PW_eq}. Note that droplets in the first and third solutions walk slightly slower than a single droplet while the second and fourth walk slightly faster.%

\subsection{Linear stability analysis}

To understand the stability of this mode, we use an approach similar to that used by \citet{Oza2013} to explore single droplet walking. The linear stability analysis of parallel walkers with varying phase based on empirical observations has been performed by \citet{PhysRevFluids.3.013604}. Consider a perturbation to the equilibrium solution $\mathbf{r}_{1}=(\epsilon x_{11}(t)H(t),ut+\epsilon y_{11}(t)H(t))$ and $\mathbf{r}_{2}=(d+\epsilon x_{21}(t)H(t),ut+\epsilon y_{21}(t)H(t))$, with the Heaviside step function $H(\cdot)$ included to introduce the perturbation at $t=0$. Substituting this form into equation \eqref{eq_1}, linearizing and taking Laplace transforms of the resulting equations, we obtain the matrix equation
\begin{equation*}
\mathsf{A}_{PW}(s)\mathbf{X}(s)=\mathbf{X}_0(s)
\end{equation*}
where
\begin{gather*}
  \mathbf{X}(s)=
\begin{bmatrix}
    X_{11}(s)\\
    Y_{11}(s)\\
    X_{21}(s)\\
    Y_{21}(s)\\
\end{bmatrix}=
\mathscr{L}\begin{bmatrix}
    x_{11}(t)\\
    y_{11}(t)\\
    x_{21}(t)\\
    y_{21}(t)\\
\end{bmatrix}
,\\[8pt]
\mathbf{X}_0(s)= (s+1)\mathbf{x}(0)+\dot{\mathbf{x}}(0),
\end{gather*}
and
\begin{equation*}
\mathsf{A}_{PW}(s)=(\kappa s^2 + s)\mathsf{I}_4+\beta \mathsf{K}
\end{equation*}
where $\mathsf{I}_4$ is the $4\times 4$ identity matrix and
\begin{widetext}
\begin{align*}
\mathsf{K}&= \mathscr{L}{  
\begin{bmatrix}
     p_1(u,0,t) & 0 & f^1_0(u,d,t;d^2) & -f^0_1(u,d,t;d) \\
     0 & f^1_2(u,0,t;1) & -f^0_1(u,d,t;d) & f^1_2(u,d,t;1)\\
f^1_0(u,d,t;d^2) & f^0_1(u,d,t;d) & p_1(u,0,t) & 0 \\
f^0_1(u,d,t;d) & f^1_2(u,d,t;1) & 0 & f^1_2(u,0,t;1) \\
\end{bmatrix}
}\notag\\[12pt]
&\quad -\int_{0}^{\infty} 
\begingroup 
\setlength\arraycolsep{0pt}
\begin{bmatrix}
     p_1(u,0,z)+f^0_0(u,d,z;d^2) & -f^0_1(u,d,z;d) & 0 & 0 \\
     -f^0_1(u,d,z;d) & f^1_2(u,0,z;1)+q_2(u,d,z) & 0 & 0\\
0 & 0 & p_1(u,0,z)+f^0_0(u,d,z;d^2) & f^0_1(u,d,z;d) \\
0 & 0 & f^0_1(u,d,z;d) & f^1_2(u,0,z;1)+q_2(u,d,z) \\
\end{bmatrix} 
\endgroup
\text{d}z
\end{align*}
\end{widetext}
Here
\begin{gather*}
p_m(u,d,z)=m\frac{\text{J}_1(\sqrt{u^2z^2+d^2})}{\sqrt{u^2z^2+d^2}} \text{e}^{-z}, \\
q_n(u,d,z)=\frac{(uz)^n}{\sqrt{u^2z^2+d^2}} \left( \frac{\text{J}_1(\sqrt{u^2z^2+d^2})}{\sqrt{u^2z^2+d^2}} \right)'\text{e}^{-z}\\
\text{and} \quad f^m_n(u,d,z;\sigma)=p_m(u,d,z)+\sigma q_n(u,d,z).
\end{gather*}

\begin{figure}
\hspace*{-1.5cm} 
\centering
\includegraphics[width=7.5cm]{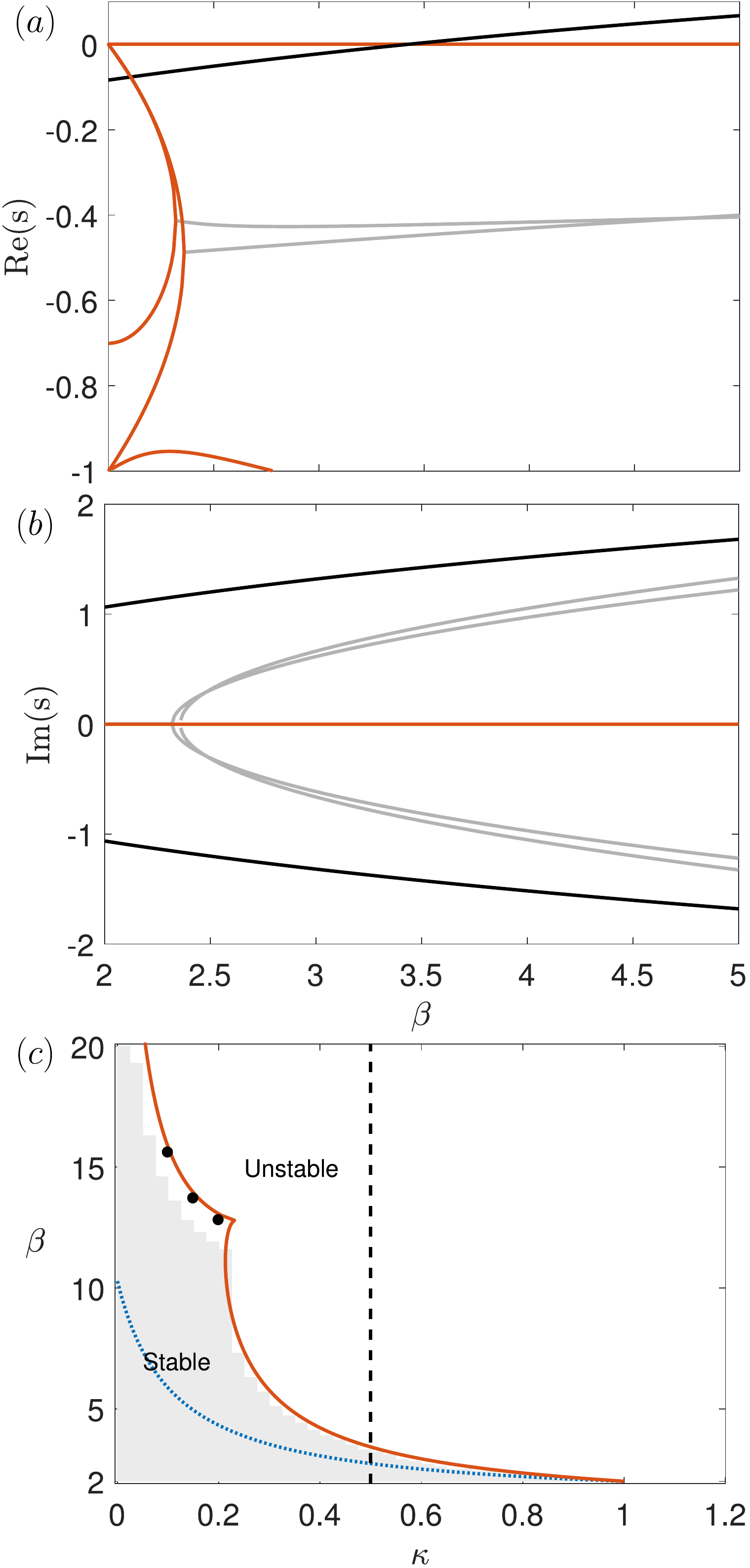}
\caption{Parallel walkers: (a) Real and (b) Imaginary part of the poles as a function of $\beta$ at $\kappa=0.5$ for the first parallel walking solution ($\dpw_1(\beta)$,$\upw_1(\beta)$). Red lines show purely real poles. The black line indicates the pole (complex conjugate) which first crosses Re$(s)=0$ resulting into the bifurcation from parallel walkers to oscillating walkers. The gray lines indicate other complex conjugate poles. (c) Stability diagram of parallel walkers in the $\beta$-$\kappa$ parameter space for the first parallel walking solution $d=\dpw_1(\beta)$ and $u=\upw_1(\beta)$. The red curve divides the parameter space into a stable region (above and to the right) and an unstable region. The vertical dashed line corresponds to $\kappa=0.5$. The shading indicates the region where parallel walkers are observed in Fig.~\ref{fig:parameter_space} where $\Delta t=2^{-6}$ and the black filled circles indicate the same boundary using a reduced time step $\Delta t=2^{-8}$ for $\kappa=0.1$, $0.15$ and $0.2$ showing that it nearly coincides with the analytical bifurcation curve (red curve). The blue dotted curve indicates the bifurcation from stationary states to inline oscillations from Fig.~\ref{fig:eig_ss}(c).}
\label{pw_poles_d}
\end{figure}

The growth rates of this linear stability problem correspond to the poles of $\mathbf{X}(s)$. The functions $p^n(u,d,z)$ and $q^n(u,d,z)$ decay exponentially as $z \to \infty$, and so all the functions in the matrix equation above are analytic in the region $Re(s)\geq 0$. Hence finding the growth rates reduces to determining the roots of $\det(\mathsf{A}_{PW}(s))=0$. This was done using by simultaneously solving real and imaginary parts of $\det(\mathsf{A}_{PW}(s))=0$ using a generalized version of the modified Secant method. We find that the distances $\dpw_2(\beta)$ and $\dpw_4(\beta)$ are always unstable while $\dpw_1(\beta)$ and $\dpw_3(\beta)$ are stable for a range of $\beta$ and $\kappa$ values. Figs.~\ref{pw_poles_d}(a,b) show the real and imaginary part of the numerically calculated poles as $\beta$ varies for $\kappa=0.5$ for droplets a distance $\dpw_1(\beta)$ apart. The first mode to become unstable is a complex conjugate pair indicating an oscillatory mode emerges at the bifurcation. Note that the zero eigenvalue reflects the invariant properties of the base state. 

Fig.~\ref{pw_poles_d}(c) shows the stability diagram for parallel walkers in the $\beta$-$\kappa$ parameter space at a distance $\dpw_1(\beta)$ apart.  
The state is stable for a large window of $\beta$ when $\kappa$ is small, with the $\beta$ window reducing as the inertia $\kappa$ increases.  The stable region corresponds well with the region where parallel walkers are observed in simulations, suggesting that the bifurcations away from parallel walking are supercritical.

Note that different modes are the first to become unstable across the two stability curves shown that meet at $\kappa\approx0.23$. For $\kappa \gtrsim 0.23$ (lower curve), simulations suggest the bifurcation results in oscillating walkers as shown in Fig.~\ref{fig:parameter_space}, while for $\kappa \lesssim 0.23$ back-and-forth walkers are observed that often become unbound in simulations.


\normalsize

\section{Oscillating walkers}\label{sec:ow}

Parallel walkers bifurcate into oscillating walkers, as observed in the parameter space plot in Fig.~\ref{fig:parameter_space}. In this mode, the droplets oscillate towards and away from one another while walking. This state has been observed experimentally.\citep{Borghesi2014,PhysRevFluids.3.013604}%

\begin{figure}
\centering
\includegraphics[width=0.9\columnwidth]{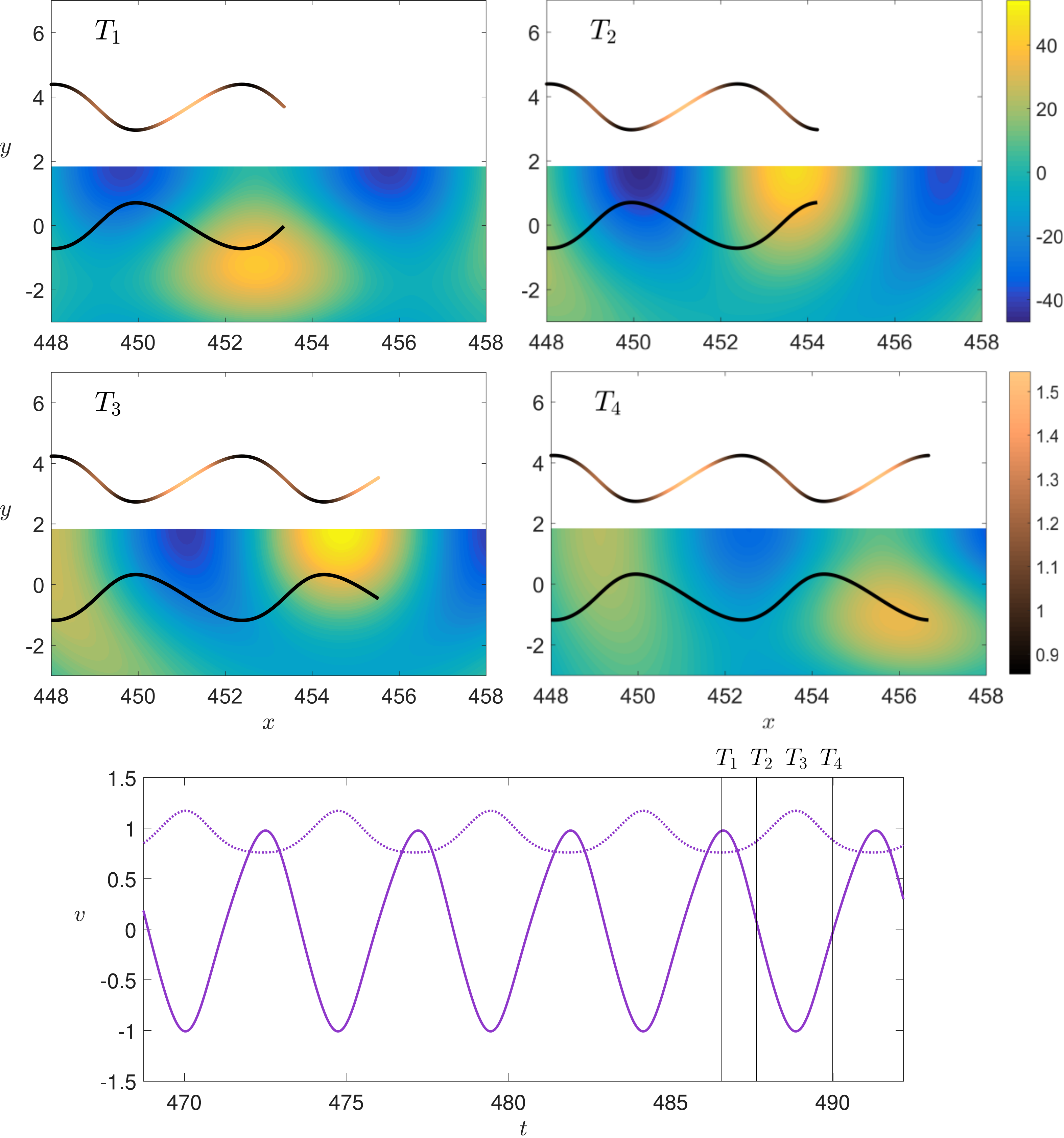}
\caption{Oscillating walkers: Droplet locations (curves), walking speed (shading on the curve in the upper half plane) and wave field (color maps in the lower half plane) at the instant of minimum forward velocity ($T_1$), an intermediate time ($T_2$), the instant of maximum forward velocity ($T_3$) and a final intermediate time ($T_4$) for $\beta=3.6$ and $\kappa=0.5$.  The bottom panel shows the inline (solid curve) and transverse (dotted curve) velocities of the droplet in the lower half plane.}
\label{pw_poles}
\end{figure}

The first mode to appear has symmetric motion of the droplets relative to the trajectory of their center of mass.  In Fig. \ref{pw_poles}, we plot the numerically simulated trajectory of such walkers, along with the underlying wave field, near the bifurcation from parallel walking. When the droplets are relatively far apart, the wave field of each droplet is discernible. However, when the droplets approach each other, their combined wave field generates a wave barrier. Note that the oscillations are primarily in the inline direction, along the direction between the two droplets, although a small oscillation also appears in the transverse, walking direction.  These two components of the oscillations are completely out of phase.  
As $\beta$ is increased for fixed $\kappa$ for these walkers, the amplitude of the oscillations grows until a new, lopsided oscillating mode appears, as described below in Section~\ref{sec:lopsided}.

Oscillating walkers also appear for $\beta$ values immediately above the tongue of unbound states at moderately small $\beta$ and moderately large $\kappa$ where inertia is too large for the droplets to be contained by the relatively weak wave field.  These are similar in structure to those bifurcating from parallel walkers, except that as $\beta$ increases their amplitude continues to increase, their inertia is sufficient to overcome the central wave barrier and they begin interchanging positions, as described below in Section~\ref{sec:switching}.

\subsection{Lopsided walkers}\label{sec:lopsided}

\begin{figure*}
\centering
\includegraphics[width=2\columnwidth]{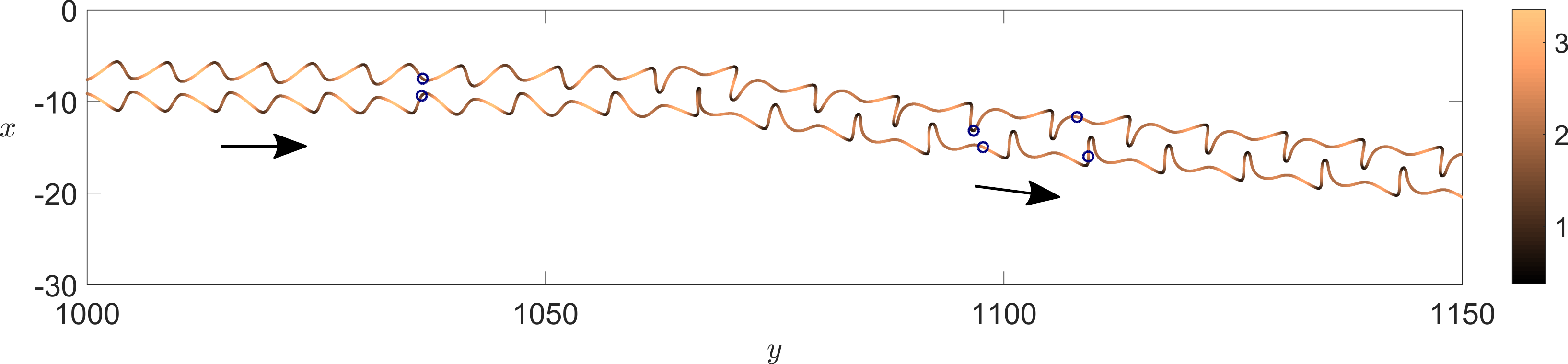}
\caption{Lopsided oscillating walkers: Trajectory for lopsided oscillating walkers for $\kappa=0.4$ and $\beta=6.7$.  The colorbar showing the speed of the droplets at a given location on the trajectory. In this simulation, the oscillating walker began in a symmetrically oscillating mode and made a sharp turn on emergence of asymmetric oscillations. Circles show the position of the droplets at a few different instances in time.}
\label{st}
\end{figure*}

Bifurcations from symmetrically oscillating parallel walkers at moderately small $\beta$ are to an asymmetrically oscillating mode as shown in Fig.~\ref{st}.  These asymmetries can be pronounced as shown, or can be more subtle with standard oscillations that are no longer perpendicular to the direction of motion.  In all cases, the center of mass of the two-droplet system now also oscillates. Where these modes are observed, they switch from an initial symmetrically oscillating state.  Except at the $\beta$ value where this mode is first observed, this switch is accompanied by an abrupt change in average direction of walking.  This abrupt change in direction is a pre-cursor to discrete-turning walkers described in Section~\ref{sec:dtw}.

\begin{figure*}
\centering
\includegraphics[width=2\columnwidth]{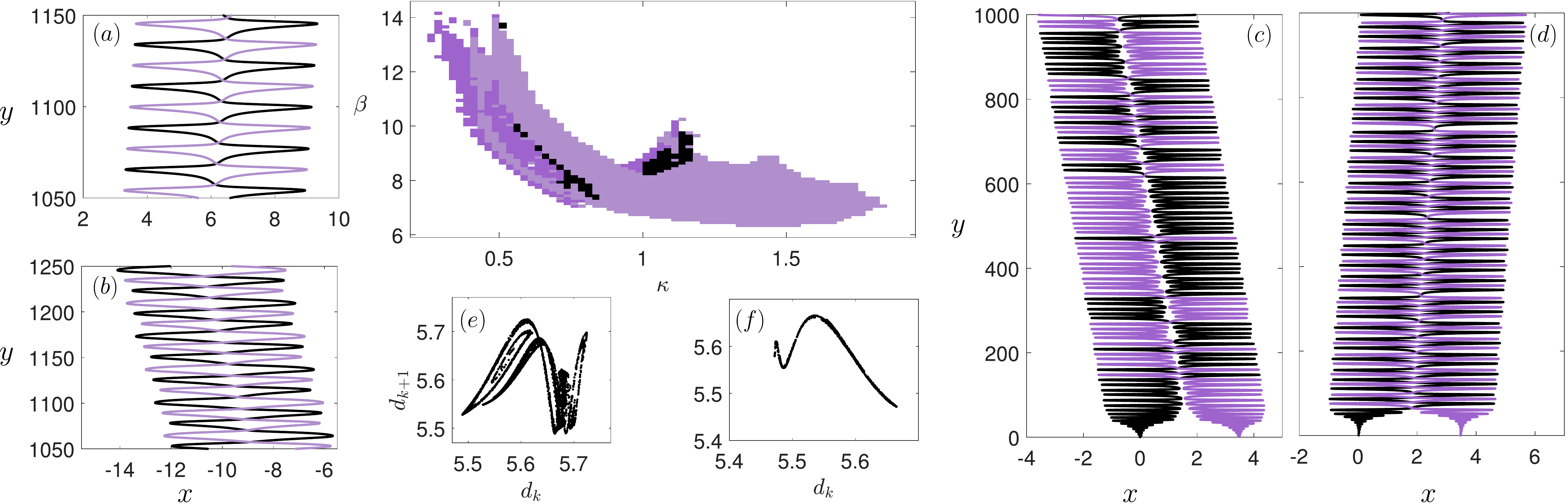}
\caption{Switching walkers: (center panel) Parameter space diagram for switching walkers indicating periodic switching and periodic amplitude changes (light purple), periodic switching and chaotic amplitude changes (black) and chaotic switching and chaotic amplitude changes (purple). Representative trajectory plots show (a) periodic switching with constant amplitude ($\kappa=0.5$ and $\beta=9.4$), (b) periodic switching with amplitude changes repeating every $3$ oscillations ($\kappa=1.2$ and $\beta=9.7$),  (c) chaotic amplitude modulations and chaotic switching ($\kappa=0.5$ and $\beta=9.1$) and (d) chaotic amplitude modulation and regular switching ($\kappa=0.575$ and $\beta=9.9)$. (e,f) First return map of the maximum distance $d_{k+1}$ in the $k+1$st oscillation as a function of the maximum distance $d_k$ in the $k$th oscillation for trajectories (c,d). The map is single-valued for trajectories with chaos only in the amplitude, while it is multi-valued for trajectories with chaos in both amplitude and switching.}
\label{switching}
\end{figure*}

\subsection{Switching walkers}\label{sec:switching}
In a tongue of parameter space in the range $7 \lesssim \beta \lesssim 12$ and $0.4 \lesssim \kappa \lesssim 1.8$, switching walkers are observed. These are symmetrically oscillating walkers whose amplitude is sufficient  to result in the droplets interchanging position.

%
Intertwined regions of periodic and chaotic switching are found as shown in Fig.~\ref{switching}.  There are two main types of periodic switching: In the first, the amplitude of oscillations is constant and switching taking place periodically. In the second, the amplitude changes periodically in addition to the switching.  Typical trajectories for each type are shown in Figs.~\ref{switching}(a,b).  In most of the periodic switching trajectories, the droplets switch after every oscillation (called period-$1$ switching) although higher-period switching is also observed.

Switching of droplets also occurs in a chaotic fashion. The chaos can either be just in the amplitude with regular switching or in both the amplitude and the switching of the oscillating walkers as shown in Fig.~\ref{switching}(c,d). Figs.~\ref{switching}(e,f) show the first return map of the maximum distance $d_{k+1}$ as a function of $d_k$ for the chaotic trajectories shown.  The return map is multi-valued when there is chaos in both amplitude and switching, while it is single valued for the case when there is chaos only in the amplitude.  Moreover, the former seems to show hints of stretching and folding similar to a Smale horseshoe map.\cite{guckenheimer2013nonlinear} 

We emphasise that such modes are unphysical because the two droplets occupy the same location as they cross their center line.  To correct this, the interaction between two nearly touching droplets would need to be included in the governing model.  This is beyond our present scope.

\section{Wandering walkers}\label{ed}

More exotic behaviors are observed in the simulations once they begin to deviate from on-average straight-line walking.  A detailed analysis of these is beyond our scope, but here we describe some of the more interesting dynamical features.

\subsection{Back-and-forth walkers}
\begin{figure}
\centering
\includegraphics[width=\columnwidth]{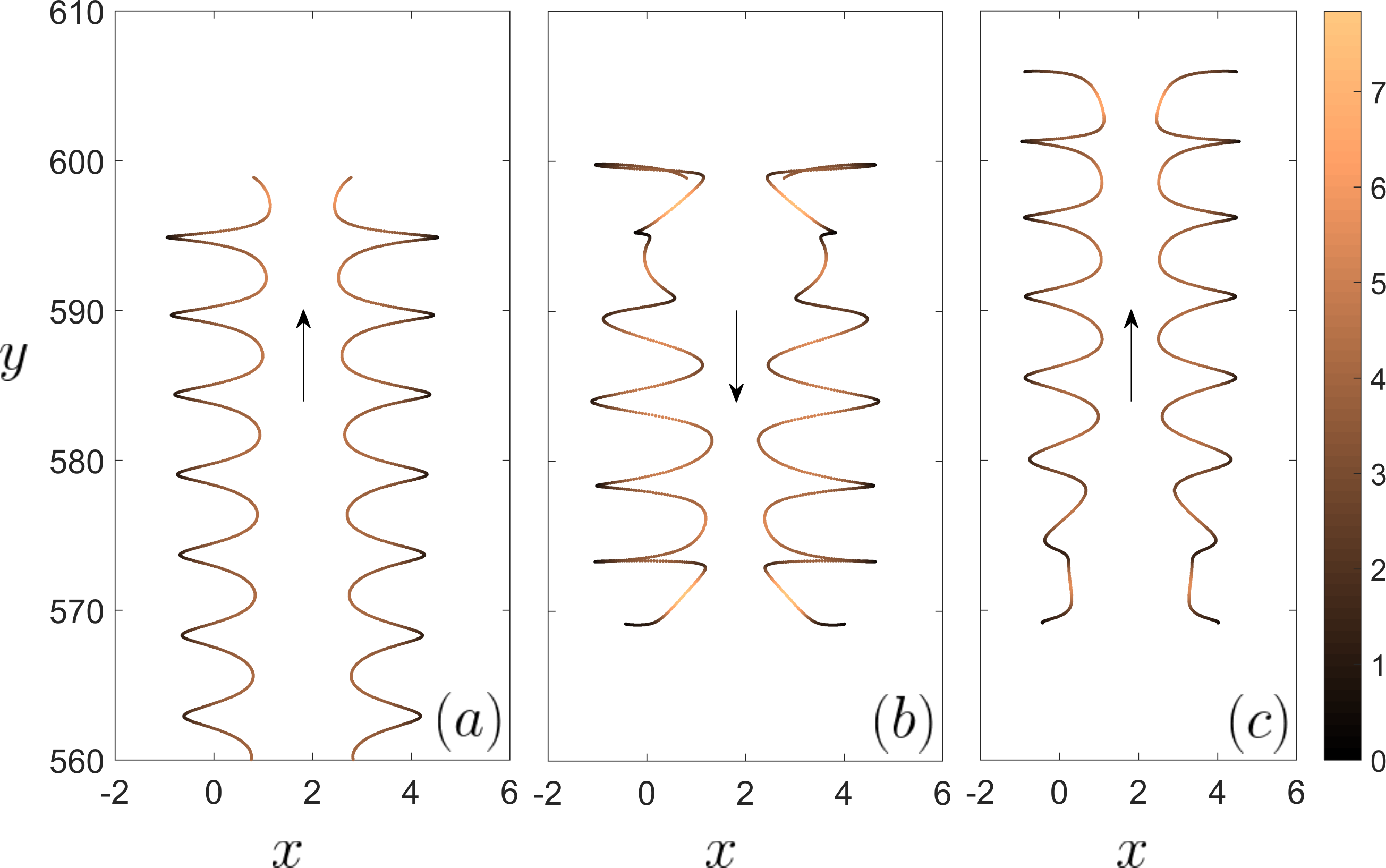}
\caption{Back-and-forth walkers: Trajectory of back-and-forth oscillating walkers at $\kappa=0.1$ and $\beta=15.1$. (a) Initial motion for $0<t\lesssim 210.9$, (b) the pair reverse their direction for $210.9 \lesssim t \lesssim 226.5$ and (c) reverse it again for $226.5\lesssim t\lesssim 242.2$. The colorbar indicates the speed of the droplets.}
\label{bfw}
\end{figure}

These are rare states found for small inertia, $\kappa<0.25$, and occur shortly after parallel walkers become unstable. The droplets in these trajectories walk as oscillating walkers but they reverse their direction of walking after several oscillations (Fig.~\ref{bfw}). This type of dynamics seems to be unstable and although observed at intermediate times in most of the simulations in this region of parameter space, the droplets usually unbind before the end of the simulation.


\begin{figure}
\centering
\includegraphics[width=\columnwidth]{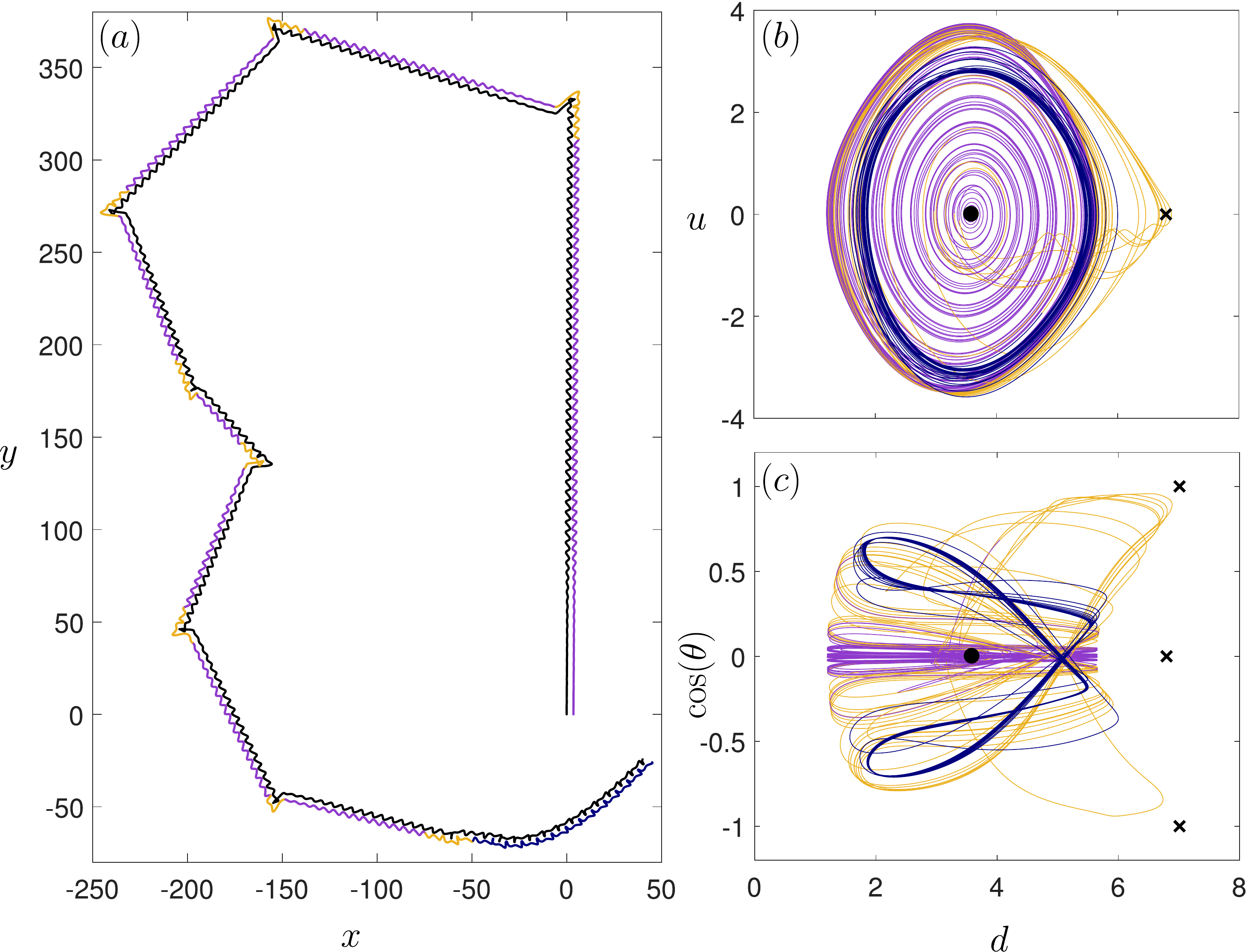}
\caption{Emergence of discrete turning walkers: (a) Trajectory at $\kappa=0.5$ and $\beta=5.3$ showing symmetrically oscillating walkers (purple) making multiple discrete turns and eventually settling into a lopsided mode (navy blue). The transient behavior during the discrete turns is shown as yellow. Projections of phase space dynamics in (a) $u$-$d$ and (b) $\cos(\theta)$-$d$ plane where $u$ is the speed of the droplet in the direction of the line joining the droplets, $d$ is the distance between the droplets and $\cos(\theta)$ is the cosine of the angle between the velocity of the center of mass and the line joining the droplets. Relevant equilibria of the $\dpw_1$ (black filled circle) and $\dpw_2$ (black cross) parallel walking modes and the $\dcm_2$ (black cross) chasing mode (see Appendix~\ref{chase}) are indicated.}
\label{phase}
\end{figure}

\subsection{Discrete-turning walkers}\label{sec:dtw}

The bifurcations from parallel walkers to (symmetrically) oscillating walkers to lopsided walkers culminate in discrete-turning walkers in a narrow region near $0.4\lesssim \kappa \lesssim 0.6$ and $3\lesssim \beta \lesssim 8$.  In this regime, the two droplets perform repeated quantized turns after walking in an on-average straight line for some distance. Fig.~\ref{phase}(a) shows a typical trajectory. In Figs.~\ref{phase}(b,c), we show two phase-space projections illustrating the lead-up to discrete-turning walkers with variables of the distance $d$ between the two droplets, the speed $u$ of the droplets in the direction of the line joining them and $\cos(\theta)$ the cosine of the angle between the velocity of the center of mass and the line joining the droplets. Relevant equilibria of the $\dpw_1$ and $\dpw_2$ parallel walking modes and the $\dcm_2$ chasing mode (see Appendix~\ref{chase}) are indicated as black filled circles and crosses.  In the trajectory shown in Figs.~\ref{phase}(a), the droplets start out as symmetrically oscillating walkers and make multiple discrete turns before settling into a stable lopsided motion. The limit cycle associated with walkers oscillating symmetrically around the $\dpw_1$ equilibrium is shown by the purple curves and the stable lopsided walkers are shown by the navy blue curves. Turns are shown in yellow. When a pair of symmetrically oscillating walkers attempts to transition from the symmetric mode to the lopsided, it gets flung towards the chasing fixed point in the phase space as shown in Fig.~\ref{phase}(c). This fixed point being unstable, brings the droplets back to the symmetrically oscillating walkers mode at $\dpw_1$. This loop near the chasing fixed point in the phase space corresponds to the actual turn in the trajectory.

\begin{figure}
\centering
\includegraphics[width=0.8\columnwidth]{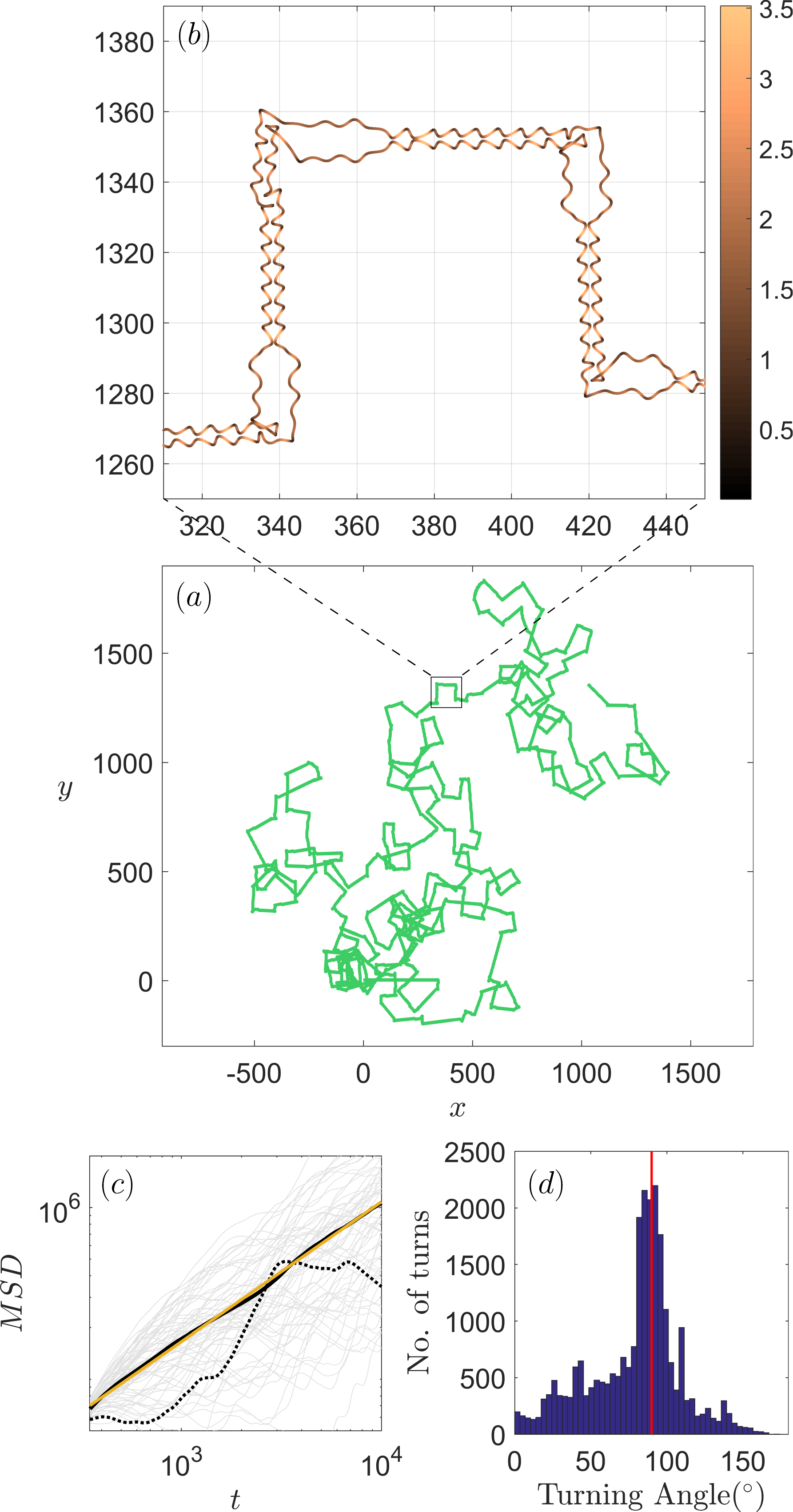}
\caption{Right-angled discrete-turning walkers: (a) Trajectory at $\kappa=0.5$ and $\beta=6$ indicating random walk-like behavior. (b) Focusing on individual turns indicates that oscillating walkers at $\dpw_1$ temporarily go to $\dpw_3$ before cascading back to $\dpw_1$. The colorbar indicates the speed of the droplets. (c) Mean squared displacement (MSD) versus time: individual trajectories are shown as light gray curves, the trajectory in (a) is shown as the black dotted curve and the ensemble average over $160$ simulations (at $\kappa=0.5$ and $\beta=6$ with noise in initial conditions) is shown as the solid black curve. Curve fitting suggests the diffusion exponent is $0.815\pm 0.002$ (solid yellow line), indicating subdiffusive behaviour. (d) Distribution of turning angles from the ensemble of simulations indicates a strong peak near $90^{\circ}$ (red vertical line).}
\label{rt}
\end{figure}

As $\beta$ is increased, it appears that the loops towards the chasing mode begin to dominate and the two droplets briefly chase one another before decaying either to the $\dpw_1$ parallel walking fixed point or the $\dpw_3$ fixed point. When it goes to the $\dpw_3$ parallel walking fixed point, it is accompanied by a turn which is nearly right angled and then cascades back to the parallel walking distance $\dpw_1$ as symmetrically oscillating walkers. We call these right-angled discrete turning walkers and they are shown in Fig.~\ref{rt}.

The underlying wave field shows that the turns are due to one of the walkers being reflected from a wave barrier.  On studying the statistics of the turning angles, we find a strong peak near $90^\circ$ (Fig.~\ref{rt}(d)), which is also evident from the trajectories.  Nearly right-angle turns are observed for all simulations in this region.  At larger scales, the trajectory appears like a random walk (Fig.~\ref{rt}(a)). By calculating the mean squared displacement as a function of time for an ensemble of simulations at $\kappa=0.5$ and $\beta=6$, we find a sub-diffusive exponent of $0.815 \pm 0.002$. Such discrete turning behavior has been observed for a single floating water droplet on the surface of a vertically vibrated highly viscous silicone oil bath.\citep{Ebata2015} 

A region of discrete-turning walkers is also observed in a small window at $\kappa\approx 0.4$ and $\beta\approx 11$.  These behave similarly, except they tend to walk in straight lines for longer before abruptly turning.  A larger region of discrete-turning walkers is observed between $1\lesssim \kappa \lesssim 2.5$ and $\beta \gtrsim 7$.  These are switching walkers and a typical trajectory is shown in Fig.~\ref{ww}(a).  Note that in these trajectories, the droplets do not necessarily occupy the same location at the same time.

Discrete-turning walkers are not always stable. When they are unstable, they typically unbind (as indicated in the mixed gray/green region in Fig.~\ref{fig:parameter_space}), although occasionally they cascade into tight orbits. These are reminiscent of cascades from oscillating walkers to orbits that have recently been observed in experiments.\citep{PhysRevFluids.3.013604}

\subsection{Continuously turning walkers}

\begin{figure*}
\centering
\includegraphics[width=2\columnwidth]{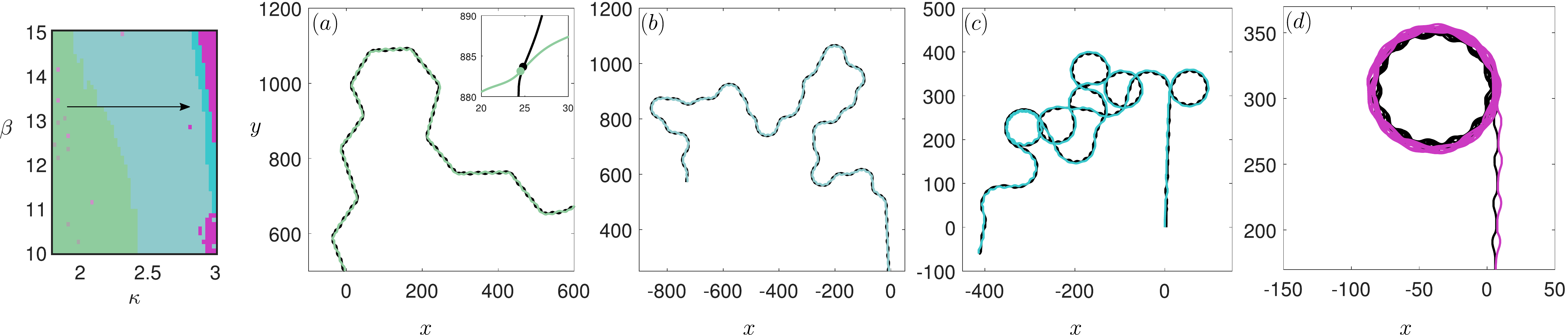}
\caption{Wandering walkers: $\beta$-$\kappa$ parameter space plot along with a progression of trajectories at fixed $\beta=13$ showing the transition from discrete-turning walkers to closed trajectories. (a) Discrete-turning walkers at $\kappa=1.9$ become (b) continuously turning walkers at $\kappa=2.5$ leading to (c) circular loops at $\kappa=2.825$ and eventually (d) stable circular trajectories at $\kappa=3$. The droplets are not always side-by-side during this wandering motion, this is shown in inset of (a) where the droplets do not reach the crossing point simultaneously.}
\label{ww}
\end{figure*}

For larger inertia, these abrupt discrete turns become smoothed, as shown in the progression of trajectories for increasing $\kappa$ and fixed $\beta$ in Fig.~\ref{ww}. For sufficiently large inertia, the turns become a series of loops and eventually closed circles.  Note that for $\kappa\lesssim 2.5$, the droplets are switching position.  Although their trajectories cross, they do not necessarily occupy the same position at the same time.


\subsection{Closed trajectories and nearly closed trajectories}

Remarkably, we find that initially parallel walkers traveling in a straight line can ultimately settle into closed trajectories.  Such states are primarily observed at high inertia as indicated by the pink region in the parameter space plot Fig.~\ref{fig:parameter_space}.  In this region, the trajectories are circles, as  shown in a representative plot in Fig.~\ref{ww}(d).

 \begin{figure}
\centering
\includegraphics[width=\columnwidth]{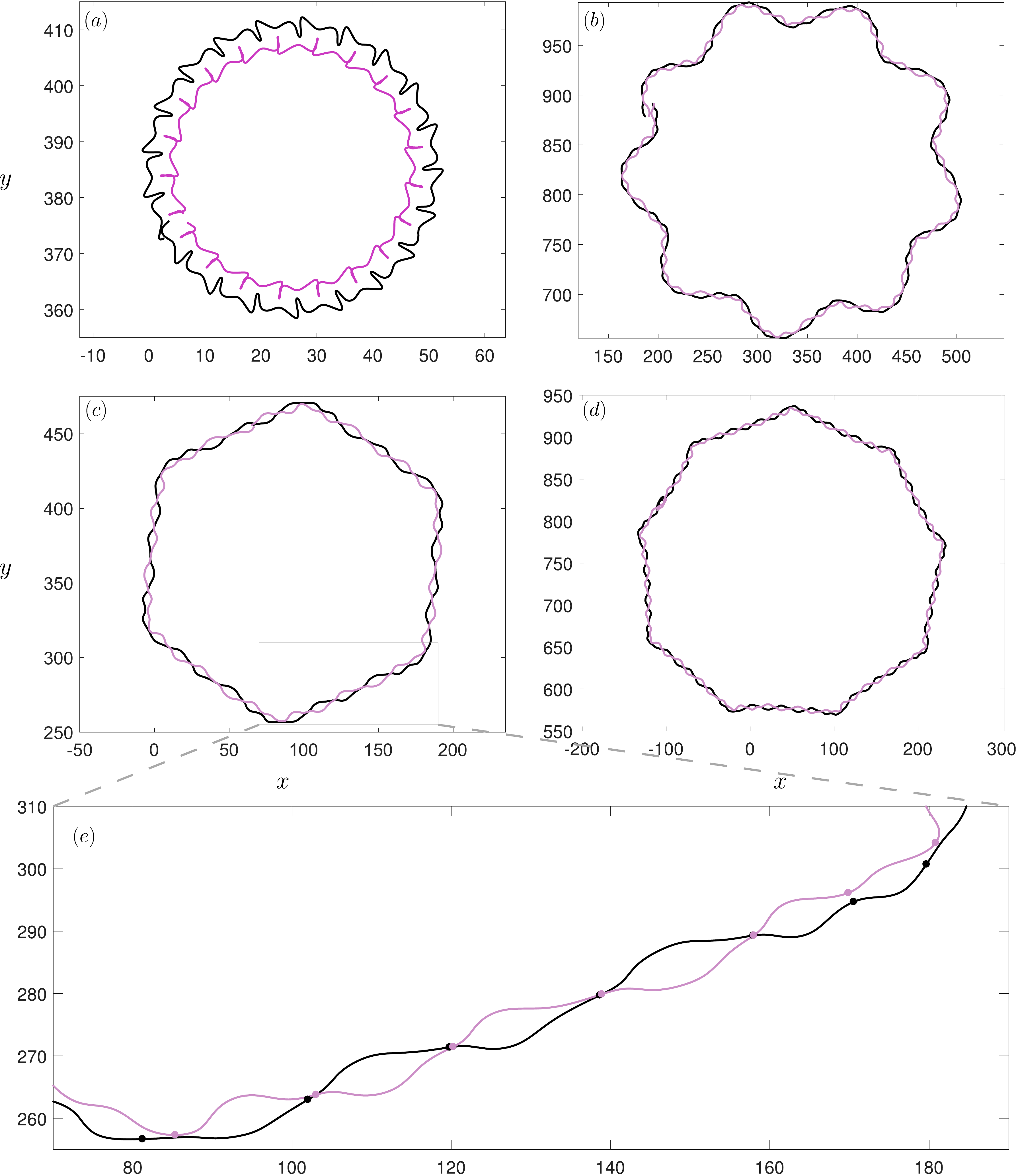}
\caption{Closed trajectories: (a) Closed circles with lopsided walkers for $\kappa=0.6$ and $\beta=4$, (b) exterior of an enneagram for $\kappa=1.875$ and $\beta=18$, (c) hexagon for $\kappa=1.85$ and $\beta=14.2$, and (d) nonagon for $\kappa=1.575$ and $\beta=16$. The polygonal structures were only traversed two or three times before the end of a simulation (only one traverse is plotted to show the structure), except for the hexagon where we have extended the simulation to $43$ traverses.  In all cases, some precession was apparent. (e) one side of the hexagon showing that the droplets are not always side-by-side and hence don't always approach the crossing point simultaneously.}
\label{ct}
\end{figure}

Circular closed trajectories also appear near $0.4\lesssim \kappa \lesssim 0.6$ and $\beta\approx 4$ or $\beta\approx 8$.  In the former region, the droplet mode is a lopsided oscillation as shown in Fig.~\ref{ct}(a), while in the latter it is a symmetric oscillation.   Rare regular polygons also appear in isolation in the parameter space, including a hexagon (Fig.~\ref{ct}(c)), an octagon (not shown) and a nonagon (Fig.~\ref{ct}(d)).  In the closed-trajectory region near $\kappa\approx 1.8$ and $\beta\gtrsim 15$, smoothed star-shaped trajectories are observed such as the exterior of an enneagram in Fig.~\ref{ct}(b). Polygons shaped orbits have been previously observed for diametrically opposed orbiting walkers\citep{PhysRevFluids.2.053601} but the closed trajectories we observe here are much larger in extent. We note that these polygonal structures are very sensitive to the numerical time step chosen, and we have not been able to reproduce them at smaller time steps although we can reliably reproduce them with different initial conditions.  We think this is because the parameters at which they form change slightly with the modified time step and we have not been able to find the exact values at which they reappear.

Intriguingly, in all closed trajectories, only the waves from the previous two oscillations of the droplets have not decayed to less than a tenth of their initial value (estimated from the location of the droplets two units of time earlier, where the exponential decay of the amplitude is $\mathrm{e}^{-2} \approx 1/10$).  In particular for the polygonal paths, this suggests that the waves from the previous turn are not directly contributing to the next turn.  However, a ``memory'' of the previous turn is retained by the system as shown in Fig.~\ref{ct}(e): the droplets are not walking symmetrically with the inner droplet on a turn leading its partner along the edges.


\section{Conclusions}\label{sec:concl}

In this paper, we have taken the Oza--Rosales--Bush stroboscopic walking-droplet model\cite{Oza2014a} as a theoretical pilot-wave description and explored the remarkable range of possible behaviors for a pair of droplets initially walking in parallel.  With increasing inertia $\kappa$ and/or wave forcing $\beta$, the droplets' motion gains degrees of freedom, commencing from a stationary pair where drag dominates both inertia and wave forcing.  The droplets first gain a single translational degree of freedom, either oscillating in place for larger $\kappa$ or parallel walking at constant speed and constant separation for larger $\beta$.  For larger $\beta$ and moderate $\kappa$, both modes are apparent and the droplets oscillate towards and away from one another with their center of mass moving in a straight line.  For larger $\beta$ still, the droplets perform this motion with random changes in direction by $180^\circ$ before gaining an additional degree of freedom with increased $\kappa$ by taking discrete turns of less than $180^\circ$ while walking.  With sufficient inertia, these turns eventually become continuous.  Surprisingly, we find that droplets only unbind if the wave forcing $\beta$ is large and inertia $\kappa$ is moderately small or in a narrow tongue where $\beta$ is small and $\kappa$ moderately large.  
For large $\beta$ and $\kappa$, the states observed at long times are intriguing: closed trajectories with effective diameters many tens of Faraday wavelengths and many times the wavelengths of the droplets' oscillations towards and away from one another.  These closed trajectories can be either regular polygons or circles.  

Our investigation has reproduced all states that have been observed experimentally: inline oscillations, parallel walkers and symmetrically oscillating walkers (promenading pairs).\cite{Borghesi2014,PhysRevFluids.3.013604}  Our simulations agree quantitatively with where oscillating walkers have been observed (white curve in Fig.~\ref{fig:parameter_space}) except at the highest memories.  Parameters for existing experimental setups are restricted to a wedge of parameter space between the white dashed curves in Fig.~\ref{fig:parameter_space}.  Besides the experimentally observed states, we predict switching modes, discrete-turning walkers and closed circular trajectories in this region.  Switching walkers are unlikely to be observed in the form described here and the droplets might either bounce off one-another, coalesce, or possibly continue walking as a condensed pair.  The important facet of an evolving impact phase in experiments may modify or even suppress any turning mode.  It would be interesting to explore whether any of the behaviors are realized.

Our investigation has focused on modes derived from parallel walkers.  Another fundamental mode is chasers, where the droplets walk one behind the other in a straight line at constant speed. These are not observed in the parameter space of Fig.~\ref{fig:parameter_space} and the linear stability results of Section~\ref{SS} suggest that parallel walking and inline modes dominate chasing. \citet{durey_milewski_2017} presented a brief study of droplet trains in their model incorporating the vertical dynamics and found that two-droplet trains (equivalent to chasers) are unstable for identical, in-phase droplets with general perturbations but can become stable for out-of-phase droplets.  We briefly explore the stability of the chasing mode for in-phase droplets using the stroboscopic model in Appendix~\ref{chase}.  
\begin{acknowledgments}
We are grateful to John Bush, Andy Hammerlindl, Joel Miller, and Tapio Simula for useful discussions. This research was partially funded by an Australian Government Research Training Program (RTP) Scholarship to R.V. 
\end{acknowledgments}

\appendix

\section{Numerical Method}\label{numerics}
We solve the trajectory equations presented in \eqref{eq_1} using a modified Euler method. For $t<0$, the droplets are assumed to be in a parallel walking state with $\mathbf{r}_{O1}=(x_{O1},y_{O1})=(0,\upw_1(\beta)t)$ and $\mathbf{r}_{O2}=(x_{O2},y_{O2})=(\dpw_1(\beta),\upw_1(\beta)t)$. The new position of the droplet is calculated form the old position using a forward Euler step as follows:
\begin{equation*}
x_i(t_{n+1})=x_i(t_n)+\Delta t u_i(t_n)
\end{equation*}
\begin{equation*}
y_i(t_{n+1})=y_i(t_n)+\Delta t v_i(t_n)
\end{equation*}

\begin{figure}
\centering
\subfigure{%
\includegraphics[width=7cm]{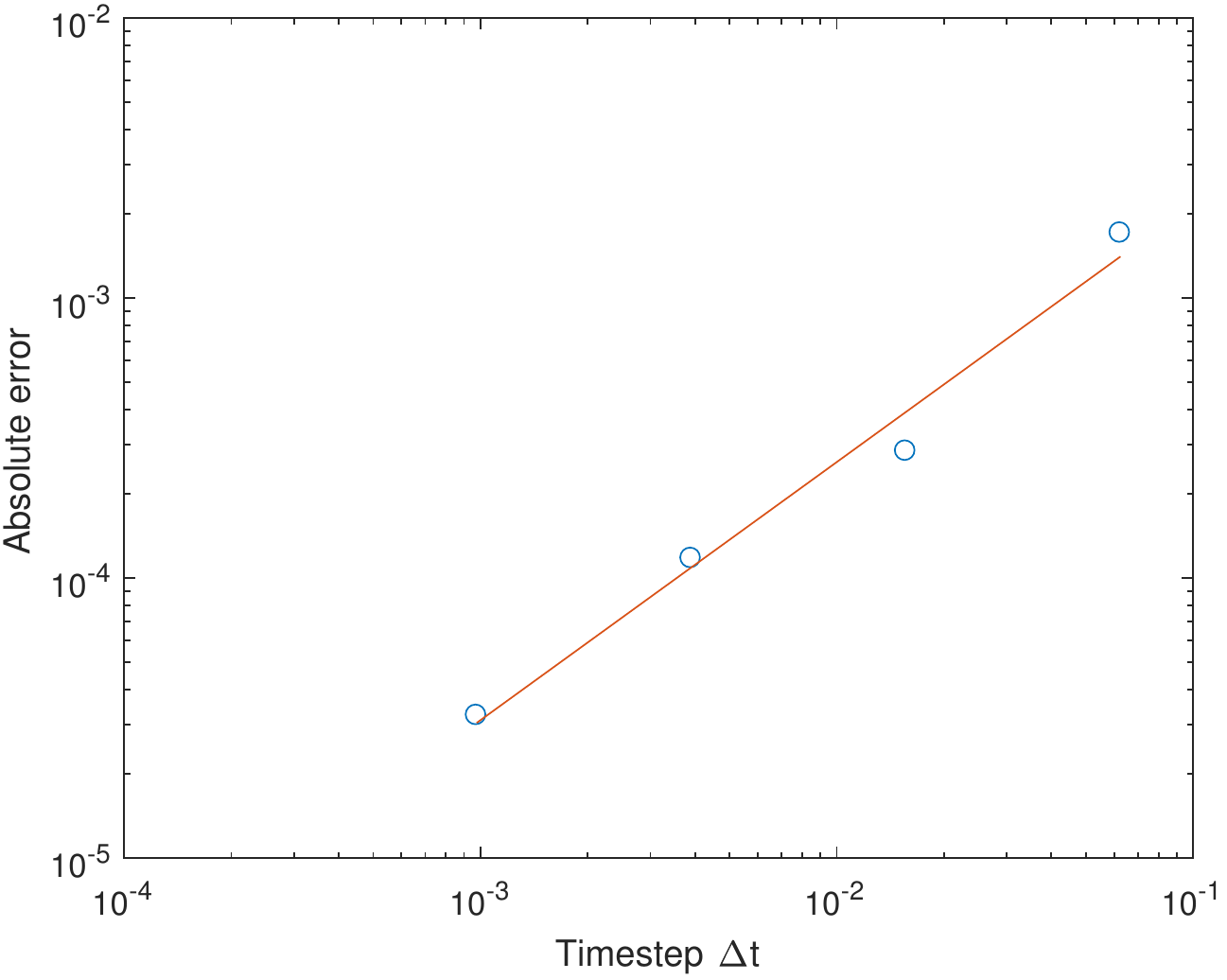}}
\quad
\caption{Comparison of Parallel Walking numerical solution using the modified Euler method with the exact solution for timesteps $\Delta t =2^{-4}$, $2^{-6}$, $2^{-8}$ and $2^{-10}$. The absolute difference in error in the parallel walking velocity is plotted for different timesteps (blue circles) with a line of best fit (orange line) of gradient $\approx 0.92$. Parameter values are $\kappa=0.5$ and $\beta=3$.}
\label{pw_c}
\end{figure}

To calculate the new velocity, we use the updated position and use a backward Euler step as follows,
\begin{widetext}
\begin{align}\label{eq:nm1}
u_i(t_{n+1})&=u_i(t_n)+\frac{\Delta t}{\kappa}\Biggl[\beta\Bigg(f_{x_{ij}}e^{-t_n}+\int_{0}^{t_n} \frac{\text{J}_1(|\mathbf{r}_i(t_{n+1})-\mathbf{r}_i(s)|)}{|\mathbf{r}_i(t_{n+1})-\mathbf{r}_i(s)|}(x_i(t_{n+1})-x_i(s))e^{-(t_{n+1}-s)}ds\notag\\
&+ \frac{\text{J}_1(|\mathbf{r}_i(t_{n+1})-\mathbf{r}_j(s)|)}{|\mathbf{r}_i(t_{n+1})-\mathbf{r}_j(s)|}(x_i(t_{n+1})-x_j(s))e^{-(t_{n+1}-s)}ds\Bigg)-u_i(t_{n+1})\Biggr]
\end{align}
\begin{align}\label{eq:nm2}
v_i(t_{n+1})&=v_i(t_n)+\frac{\Delta t}{\kappa}\Biggl[\beta\Bigg(f_{y_{ij}}e^{-t_n}+\int_{0}^{t_n} \frac{\text{J}_1(|\mathbf{r}_i(t_{n+1})-\mathbf{r}_i(s)|)}{|\mathbf{r}_i(t_{n+1})-\mathbf{r}_i(s)|}(y_i(t_{n+1})-y_i(s))e^{-(t_{n+1}-s)}ds\notag\\
&+\int_{0}^{t_n} \frac{\text{J}_1(|\mathbf{r}_i(t_{n+1})-\mathbf{r}_j(s)|)}{|\mathbf{r}_i(t_{n+1})-\mathbf{r}_j(s)|}(y_i(t_{n+1})-y_j(s))e^{-(t_{n+1}-s)}ds\Bigg)-v_i(t_{n+1})\Biggr]
\end{align}
where,
\begin{equation}\label{eq:nm3}
f_{x_{ij}}(\mathbf{r})=\int_{-\infty}^{0}\frac{\text{J}_1(|\mathbf{r}_i(t_{n})-\mathbf{r}_{Oi}(s)|)}{|\mathbf{r}_i(t_{n})-\mathbf{r}_{Oi}(s)|}(x_i(t_{n})-x_{Oi}(s))e^{s}ds+\int_{-\infty}^{0}\frac{\text{J}_1(|\mathbf{r}_i(t_{n})-\mathbf{r}_{Oj}(s)|)}{|\mathbf{r}_i(t_{n})-\mathbf{r}_{Oj}(s)|}(x_i(t_{n})-x_{Oj}(s))e^{s}ds
\end{equation}
\begin{equation}\label{eq:nm4}
f_{y_{ij}}(\mathbf{r})=\int_{-\infty}^{0}\frac{\text{J}_1(|\mathbf{r}_i(t_{n})-\mathbf{r}_{Oi}(s)|)}{|\mathbf{r}_i(t_{n})-\mathbf{r}_{Oi}(s)|}(y_i(t_{n})-y_{Oi}(s))e^{s}ds+\int_{-\infty}^{0}\frac{\text{J}_1(|\mathbf{r}_i(t_{n})-\mathbf{r}_{Oj}(s)|)}{|\mathbf{r}_i(t_{n})-\mathbf{r}_{Oj}(s)|}(y_i(t_{n})-y_{Oj}(s))e^{s}ds
\end{equation}
\end{widetext}
The integral in equations \eqref{eq:nm1} and \eqref{eq:nm2} were performed using trapezoidal rule where we consider the contribution from all the previous impacts for the first 1280 timesteps ($t=20$ using $\Delta t=2^{-6}$) and then the contribution from the last 1280 impacts for $t>20$. At 1280 previous impacts, the exponential damping factor has reached $e^{-20} \approx 10^{-9}$ so we neglect all the contribution from impacts beyond 1280 previous steps. The integral for initial condition in equations \eqref{eq:nm3} and \eqref{eq:nm4} were calculated using an adaptive Gauss-Kronrod quadrature routine built into MATLAB. The convergence of this method for the parallel walking solution is shown in Fig.~\ref{pw_c}.

Using our method, we have been able to reproduce the exotic trajectories of a single walker in a rotating frame by \citet{Tambasco2016} and \citet{Oza2014a}

\begin{figure}
\centering
\includegraphics[width=6cm]{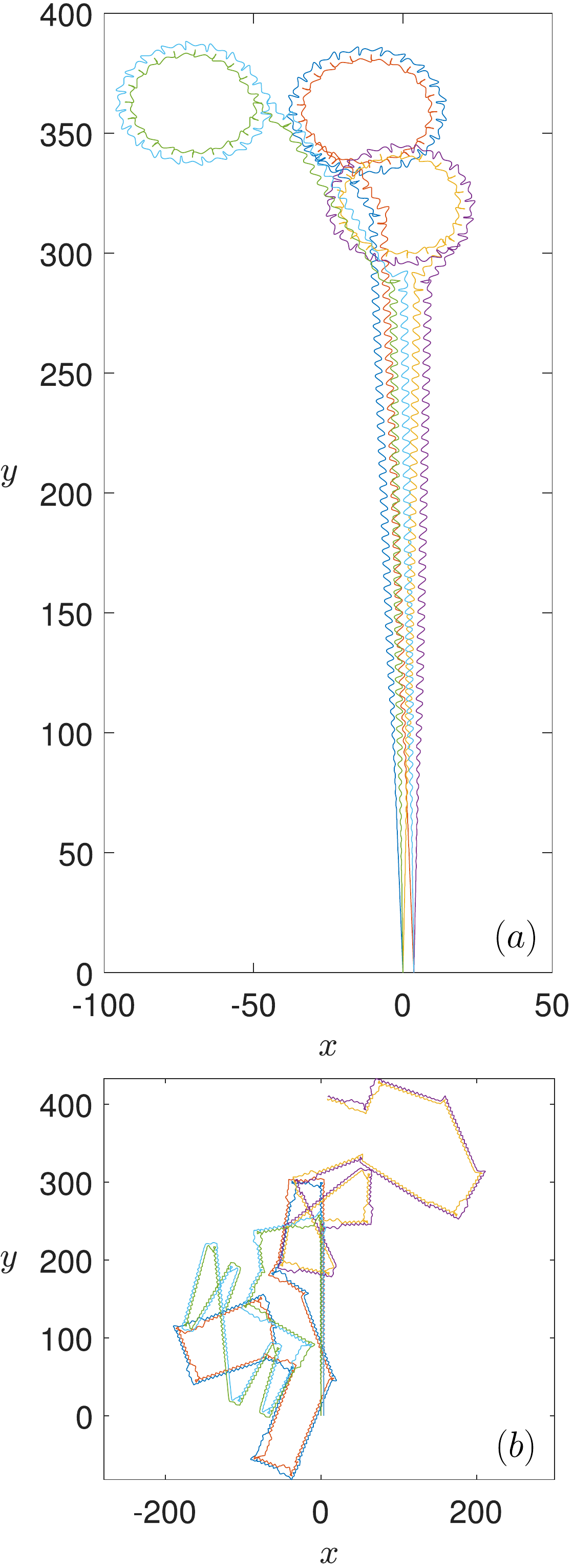}
\quad
\caption{Comparison of trajectories for (a) $\kappa=0.6$ and $\beta=4$ and (b) $\kappa=0.5$ and $\beta=6$ starting as parallel walkers with random noise using timesteps $\Delta t=2^{-6}$ (blue and orange), $2^{-8}$ (yellow and purple) and $2^{-10}$ (green and cyan). All three timesteps eventually lead to the exotic trajectory of closed circles with lopsided walkers for (a) and right-angled discrete turning walking for (b).}
\label{lp_cp}
\end{figure}

Fig.~\ref{lp_cp} shows the comparison with different timesteps of the closed circular trajectory at $\kappa=0.6$ and $\beta=4$ where the pair of walkers are in a lopsided mode and the right-angled discrete turning walkers at $\kappa-0.5$ and $\beta=6$. Simulating trajectories at this parameter value with timesteps $\Delta t =2^{-6}$, $2^{-8}$ and $2^{-10}$ with noise in initial conditions confirm that these exotic behaviours are robust.

%

\section{Chasing mode}\label{chase}

Consider two in-phase droplets chasing one another in one-dimensional motion at a constant speed $u$ and maintaining a constant separation $d$: $\mathbf{r}_{1}=(ut,0)$ and $\mathbf{r}_{2}=(ut+d,0)$. Substituting these forms into \eqref{eq_1}, we obtain the following pair of equations:
\begin{equation}\label{CM_eq_1}
u  =  \beta \Biggl(  \int_{0}^{\infty}\text{J}_1(uz)  \text{e}^{-z} \text{d}z
 + \int_{0}^{\infty}\text{J}_1(uz\mp d) \text{e}^{-z} \text{d}z \Biggr).
\end{equation}
Here the first integral 
represents the force on the droplet due to its own wave field while the second integral is the force from the other droplet's wave field.  


\begin{figure}
\hspace*{-0.5cm}   
\centering
\includegraphics[width=9cm]{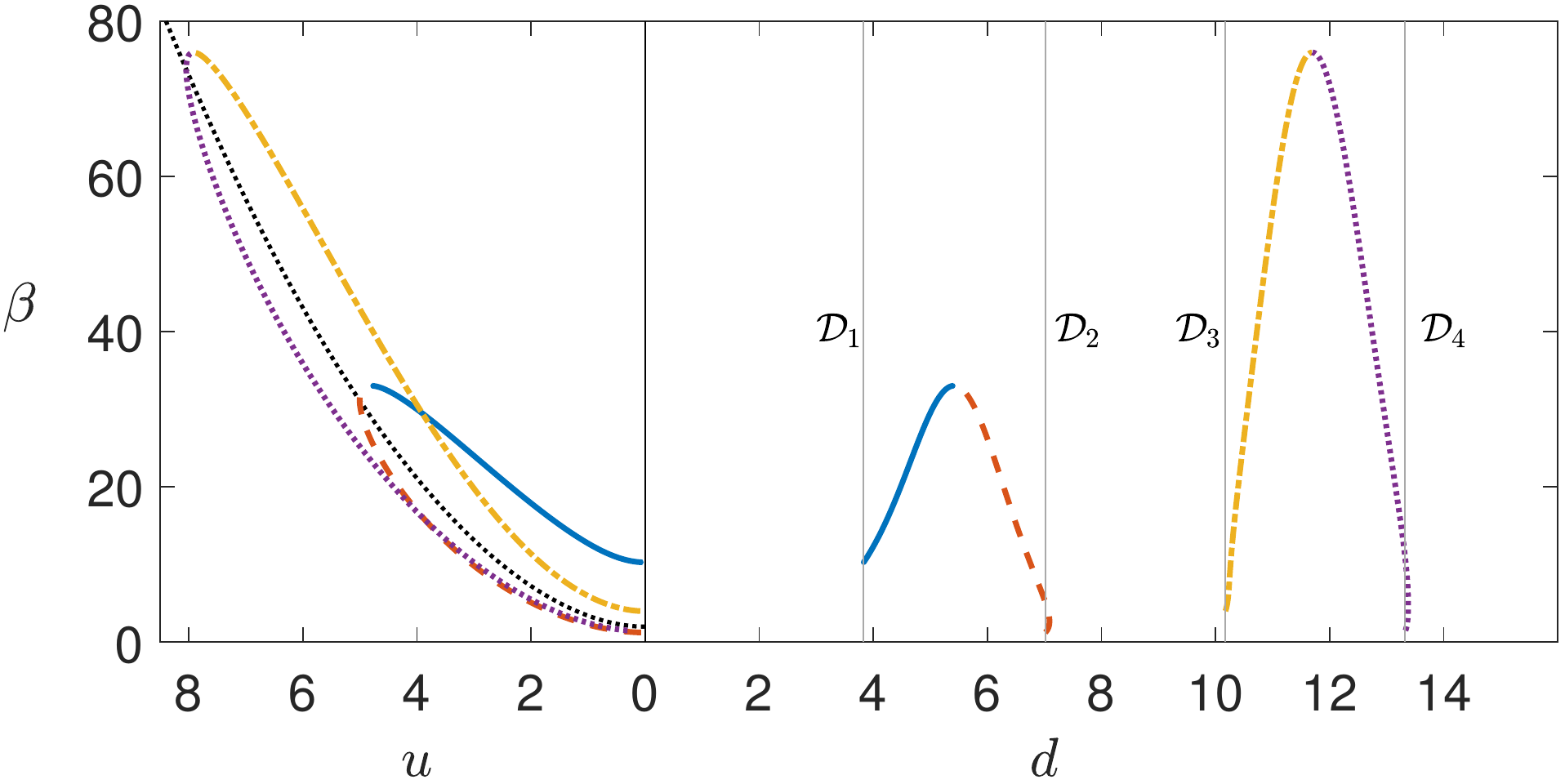}
\caption{Chasers: Equilibrium solutions $u=\ucm_n(\beta)$ and $d=\dcm_n(\beta)$ for $n=1$ (solid blue line), $2$ (dashed red line), $3$ (dashed-dotted yellow line) and $4$ (dotted purple line). Black dotted line shows the curve for a single walker.}
\label{CM_eq}
\end{figure}

Fig.~\ref{CM_eq} shows the numerical solutions of \eqref{CM_eq_1} as a function of $\beta$ (solutions are independent of $\kappa$). There are infinitely many solution pairs $u=\ucm_n(\beta)$ and $d=\dcm_n(\beta)$. Each solution pair only exists for a window of $\beta$ values. The solution first emerges from the corresponding stationary state solution at $\beta=\beta^c_n$ with $\ucm_n=0$ and $\dcm_n=\dis_n$. Pairs of solution pairs coincide and annihilate one another in a saddle-node bifurcation at the upper end of the window. 



\subsection{Linear stability analysis}\label{ls_c}

\begin{figure}
\centering
\subfigure{%
\includegraphics[width=7cm]{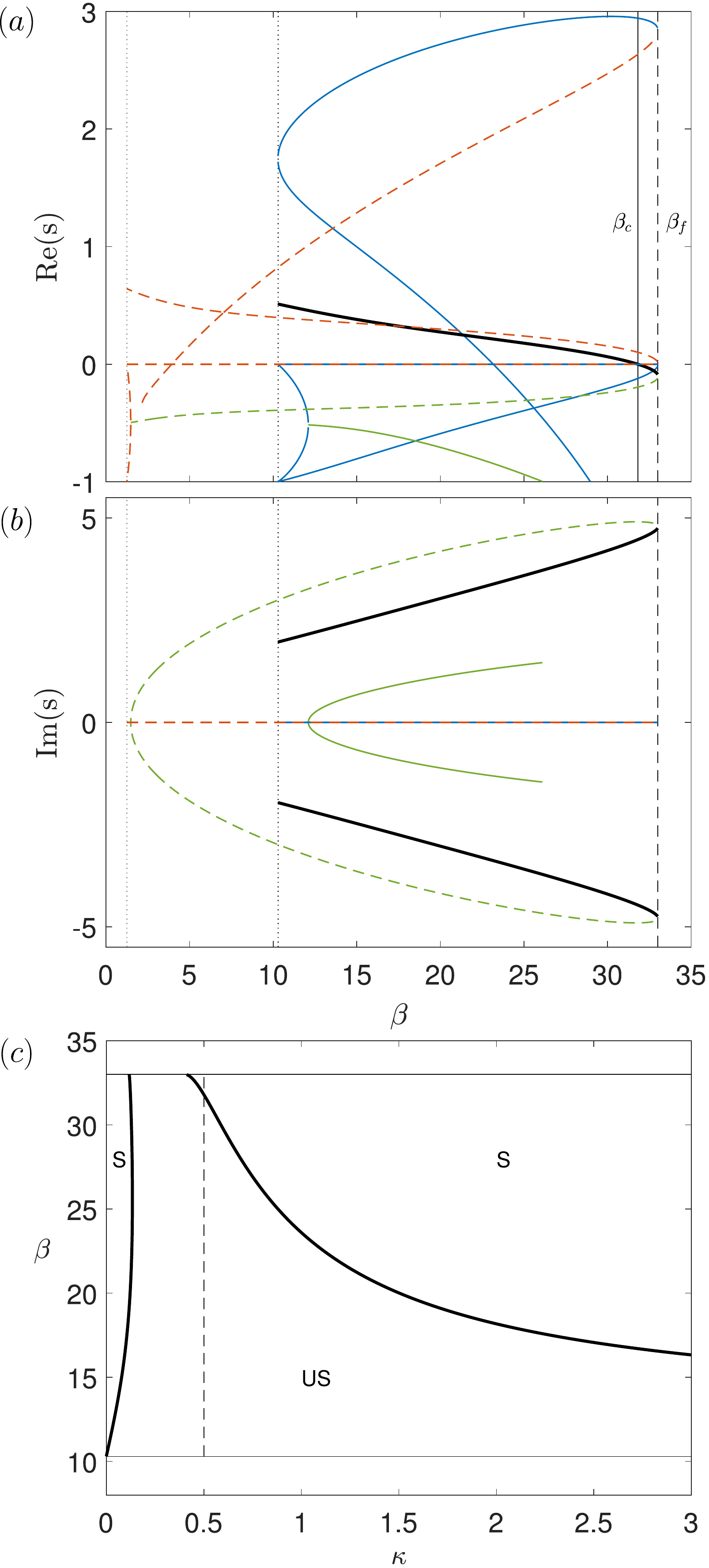}}
\quad
\caption{Chasers: 
Plot of the (a) real and (b) imaginary part of the poles as a function of $\beta$ at $\dcm_1(\beta)$ (solid lines) and $\dcm_2(\beta)$ (dashed lines) with $\kappa=0.5$. The vertical black dotted lines correspond to the start of solution at $\dcm_1(\beta)$ and $\dcm_2(\beta)$ while the vertical black dashed line corresponds to the end of the chasing solution at $\beta=\beta_f$. Poles with imaginary parts are shown in green. The thick black solid line corresponds to the pole from inline perturbation for which Re$(s)<0$ for $\beta_c<\beta<\beta_f$ where $\beta_c$ (vertical black solid line) is where it first crosses Re$(s)=0$. This pole is a complex conjugate as indicated by the thick black lines in (b) and its stability in the $\beta$-$\kappa$ parameter space is shown in (c) where the stable and unstable regions are indicated by S and US respectively. The dashed vertical line at $\kappa=0.5$ in (c) corresponds to the plots in (a) and (b).}
\label{cm_ls}
\end{figure}

To understand the stability of this mode, we consider a general perturbation to a pair of droplets in the chasing mode applied at $t=0$ as follows: $\mathbf{r}_{1}=(ut+\epsilon x_{11}(t)H(t),\epsilon y_{11}(t)H(t))$ and $\mathbf{r}_{2}=(ut+d+\epsilon x_{21}(t)H(t),\epsilon y_{21}(t)H(t))$, similar to the analysis for parallel walkers. Substituting this form into \eqref{eq_1} and linearising, we find 
\begin{widetext}
\begin{align*}
\kappa \ddot{x}_{i1} + \dot{x}_{i1} &= \beta \Biggl[\left(\int_{0}^{\infty} (\text{J}_1'(uz)+\text{J}_1'(uz\mp d))\text{e}^{-z}\text{d}z)\right)x_{i1}(t) -\int_{0}^{\infty} \text{J}_1'(uz)x_{i1}(t-z)H(t-z)\text{e}^{-z}\text{d}z\notag\\
&-\int_{0}^{\infty} \text{J}_1'(uz\mp d)x_{j1}(t-z)H(t-z)\text{e}^{-z}\text{d}z\Biggr]
\end{align*}
\begin{align*}
\kappa \ddot{y}_{i1} + \dot{y}_{i1} &= \beta \Biggl[\left(\int_{0}^{\infty} \left(\frac{\text{J}_1(uz)}{uz}+\frac{\text{J}_1(uz\mp d)}{uz\mp d}\right)\text{e}^{-z}\text{d}z\right)y_{i1}(t)-\int_{0}^{\infty} \frac{\text{J}_1(uz)}{uz}y_{i1}(t-z)H(t-z)\text{e}^{-z}\text{d}z\notag\\
&-\int_{0}^{\infty} \frac{\text{J}_1(uz\mp d)}{uz\mp d}y_{j1}(t-z)H(t-z)\text{e}^{-z}\text{d}z\Biggr]
\end{align*}
\end{widetext}
for $i=1$, $j=2$ with the negative signs, and $i=2$, $j=1$ with the positive signs.
On taking Laplace transforms on both sides, the equations can be rewritten in the matrix form
\begin{equation*}
\mathsf{A}_{\text{chase}}(s)\mathbf{X}(s)=\mathbf{X}_0(s)
\end{equation*}
where
\begin{equation*}
\mathbf{X}(s)=
\begin{bmatrix}
    X_{11}(s)\\
    Y_{11}(s)\\
    X_{21}(s)\\
    Y_{21}(s)\\
\end{bmatrix}
,\quad
\mathbf{X}_0(s)=(s+1)\mathbf{x}(0)+\mathbf{\dot{x}}(0)
\end{equation*}
and
\begin{equation*}
\mathsf{A}_{\text{chase}}(s)=(\kappa s^2 + s)\mathsf{I}_4+\beta \mathsf{K}(s).
\end{equation*}
Here $x_{i1}(t)$ are the dynamical variables in the time domain and $X_{i1}(s)$ are the dynamical variable in Laplace space and
\begin{equation*}
\mathsf{K}(s)=
\begin{bmatrix}
\text{B}(s)-\text{A}_{-} & 0 & \text{C}_{-}(s) & 0\\
0 & \text{E}(s)-\text{D}_{-} & 0 & \text{F}_{-}(s)\\
\text{C}_{+}(s) & 0 & \text{B}(s)-\text{A}_{+} & 0\\
0 & \text{F}_{+}(s) & 0 & \text{E}(s)-\text{D}_{+}\\
\end{bmatrix}
\end{equation*}
with
\begin{gather*}
\text{A}_{\mp}=\int_{0}^{\infty}\left(\text{J}_1'(uz)+\text{J}_1'(uz\mp d)\right)\text{e}^{-z} \text{d}z
,\notag\\
\text{D}_{\mp}=\int_{0}^{\infty}\left(\frac{\text{J}_1(uz)}{uz}+\frac{\text{J}_1(uz\mp d)}{uz\mp d}\right)\text{e}^{-z} \text{d}z,\\
\text{B}(s)=\mathscr{L}\left \{\text{J}_1'(ut) \text{e}^{-t}\right \}
,\quad
\text{C}_{\mp}(s)=\mathscr{L}\left \{\text{J}_1'(ut\mp d) \text{e}^{-t}\right \}
,\notag\\
\text{E}(s)=\mathscr{L}\left \{\frac{\text{J}_1(ut)}{ut} \text{e}^{-t}\right \}\:\text{and}\:
\text{F}_{\mp}(s)=\mathscr{L}\left \{\frac{\text{J}_1(ut\mp d)}{ut\mp d} \text{e}^{-t}\right \},
\end{gather*}
where $\mathscr{L}$ is the Laplace transform operator. Fig.~\ref{cm_ls}(a) and (b) shows the real Re$(s)$ and imaginary Im$(s)$ part of the poles of $\mathbf{X}(s)$ as a function of $\beta$ for chasing walkers at distance $\dcm_1(\beta)$ and $\dcm_2(\beta)$ with $\kappa=0.5$. Note that the zero eigenvalue reflects the invariant properties of the base state. There is always a transverse mode with Re$(s)>0$ for both $\dcm_1(\beta)$ and $\dcm_2(\beta)$ indicating that the chasers are always unstable for general perturbations. This hold true for all $\kappa$.  For $\dcm_1(\beta)$, we see that the only unstable mode (complex conjugate) corresponding to the inline perturbation crosses Re$(s)=0$ around $\beta_c=\beta\approx 31$ indicating that droplets are stable to inline perturbations for $\beta_c<\beta<\beta_f$, where $\beta_f$ is where the chasing solution terminates. A stability diagram in the $\beta$-$\kappa$ parameter space indicating the stable and unstable region to inline perturbations at $\dcm_1(\beta)$ is shown in Fig.~\ref{cm_ls}(c).

\section*{References}

\bibliography{PWD}

\end{document}